\documentclass[twocolumn]{aastex631}

\shorttitle{Venus Analogs from PLATO}
\shortauthors{Stephen R. Kane et al.}

\begin{document}

\title{The Expected Yield of Venus Zone Terrestrial Planets from
  PLATO}

\author[0000-0002-7084-0529]{Stephen R. Kane}
\affiliation{Department of Earth and Planetary Sciences, University of
  California, Riverside, CA 92521, USA}
\email{skane@ucr.edu}

\author[0009-0006-9233-1481]{Emma L. Miles}
\affiliation{Department of Earth and Planetary Sciences, University of
  California, Riverside, CA 92521, USA}

\author[0000-0001-7968-0309]{Colby M. Ostberg}
\affiliation{Department of Earth and Planetary Sciences, University of
  California, Riverside, CA 92521, USA}

\author[0000-0002-0000-199X]{Erika Kohler}
\affiliation{NASA Goddard Space Flight Center, 8800 Greenbelt Road,
  Greenbelt, MD 20771, USA}

\author[0000-0003-1606-5645]{James B. Garvin}
\affiliation{NASA Goddard Space Flight Center, 8800 Greenbelt Road,
  Greenbelt, MD 20771, USA}


\begin{abstract}

The characterization of terrestrial exoplanets and the conditions that
lead to divergent climate outcomes is a primary goal of exoplanetary
science. The Venus Zone (VZ) provides a framework for identifying
planets that may have experienced runaway greenhouse processes similar
to Venus, and the statistical properties of such planets bear directly
on models of planetary habitability. Here we present a quantitative
estimate of the expected yield of VZ terrestrial planets from ESA's
PLATO (PLAnetary Transits and Oscillations of stars) mission. We
combine the predicted PLATO planet yield for Earth-size
($0.8$--$1.25~R_\oplus$) and super-Earth ($1.25$--$2.0~R_\oplus$)
planets with empirical occurrence rates for VZ terrestrial planets
derived from Kepler data. Under conservative assumptions, we estimate
that PLATO will detect $\sim$170--280 VZ terrestrial planets
($0.8$--$2.0~R_\oplus$), including $\sim$40--80 Earth-size
($0.8$--$1.25~R_\oplus$) planets, during a nominal 4-year mission. For
the bright P1 sample ($V \leq 11$), we estimate $\sim$50--85
terrestrial and $\sim$13--22 Earth-size VZ detections, enabling radial
velocity mass determination and atmospheric characterization of the
most favorable targets with JWST and future facilities. We discuss the
implications of this yield for comparative studies of Earth-Venus
divergence, synergies with the DAVINCI, VERITAS, and EnVision missions
to Venus, and the role of PLATO in advancing our understanding of the
runaway greenhouse boundary.

\end{abstract}



\section{Introduction}
\label{sec:intro}

A central question in planetary science is whether Earth's clement
climate is a common outcome of terrestrial planet evolution or an
exceptional one. Venus, despite its striking similarity to Earth in
mass and radius, presents a dramatically different surface environment
characterized by temperatures exceeding 735~K and a dense
CO$_2$-dominated atmosphere with surface pressures of $\sim$93~bar
\citep{taylor2018}. Understanding how and when these two sibling
planets diverged in their evolutionary trajectories remains one of the
most important open problems in comparative planetology
\citep{kane2019d}. In the meantime, the discovery and characterization
of exoplanets has provided a statistical framework for addressing this
question. Transit surveys, including Kepler \citep{borucki2010a}, K2
\citep{howell2014}, and TESS \citep{ricker2015}, have revealed that
small planets ($R_p < 2~R_\oplus$) are abundant around FGK and M dwarf
stars \citep{fressin2013,dressing2015b,kunimoto2020b}. Particular
focus has been placed on terrestrial planets that may lie within the
Habitable Zone (HZ) of their host stars, as possible targets for
further astrobiology-related studies
\citep{kasting1993a,kane2012a,kopparapu2013a,kopparapu2014,kane2016c,hill2018,hill2023}.

A significant fraction of detected small planets receive stellar
irradiation levels that place them interior to the HZ, in a regime
where runaway greenhouse conditions may prevail. To systematically
identify planets in this regime, \citet{kane2014e} defined the Venus
Zone (VZ), the region around a star bounded on the outer edge by the
runaway greenhouse limit
\citep{kasting1993a,kopparapu2013a,kopparapu2014} and on the inner
edge by the stellar flux level at which atmospheric erosion is
expected to strip the planet of its atmosphere entirely. The transit
detection bias toward shorter-period planets means that the VZ is
particularly well-sampled by transit surveys, and \citet{kane2014e}
used Kepler data to calculate occurrence rates of VZ terrestrial
planets of $0.32^{+0.05}_{-0.07}$ for M~dwarfs and
$0.45^{+0.06}_{-0.09}$ for GK~dwarfs. These values are comparable to,
and in some cases exceed, estimates of $\eta_\oplus$, the occurrence
rate of Earth-size planets in the HZ
\citep{petigura2013a,dressing2013,bryson2021}. More recently,
\citet{ostberg2023a} presented a comprehensive demographic analysis of
terrestrial planets in the VZ, cataloging 317 planets and identifying
five high-priority targets for atmospheric characterization with the
James Webb Space Telescope (JWST). That work demonstrated the
scientific potential of exoVenus studies and highlighted the need for
larger statistical samples to constrain the frequency and diversity of
runaway greenhouse outcomes. \citet{kane2026a} further updated the
known inventory of exoVenus candidates to 370, with emphasis on
possible follow-up studies with the Habitable Worlds Observatory (HWO)
that would increase the diagnostic capabilities for post-runaway
greenhouse atmospheres
\citep{ehrenreich2012a,barstow2016a,lustigyaeger2019b,ostberg2023c}. The
atmospheric characterization of VZ planets represents a complementary
approach to the in-situ exploration of Venus by upcoming missions such
as DAVINCI \citep{garvin2022}, VERITAS \citep{cascioli2021}, and
EnVision \citep{widemann2023}, which will provide ground-truth
measurements of Venus's atmosphere, surface, and interior.

ESA's PLAnetary Transits and Oscillations of stars (PLATO) mission
\citep{rauer2014,rauer2025} is designed to detect and characterize
terrestrial planets around bright ($V \leq 13$) FGK stars, with
particular emphasis on planets in and near the HZ. With its
unprecedented combination of a large field of view
($\sim$2232~deg$^2$), multi-camera architecture (26 cameras providing
overlapping coverage), and long continuous monitoring baselines
($\geq$2~years per pointing), PLATO is poised to dramatically expand
the census of well-characterized small planets at intermediate orbital
periods. While much attention has been devoted to the PLATO yield of
Earth analogs in the HZ \citep{heller2022b,matuszewski2023},
comparatively little analysis has addressed the expected yield of
Venus analogs. The outer VZ is especially important in this regard
because it samples the regime where terrestrial planets may transition
from long-lived temperate climates to runaway greenhouse evolution.
PLATO's long baselines and bright-star sample therefore provide a rare
opportunity to populate the observational boundary between HZ and VZ
outcomes.

In this paper, we estimate the number of VZ terrestrial planets that
PLATO is expected to detect, drawing on the mission's predicted planet
yields, the properties of the PLATO stellar sample, and empirical
occurrence rates from Kepler. We assess the detectability of these
planets as a function of stellar brightness, planet size, and orbital
period, and we discuss the implications for atmospheric
characterization and comparative planetology. The structure of this
paper is as follows. Section~\ref{sec:plato} summarizes the key
specifications of the PLATO mission relevant to VZ planet detection.
Section~\ref{sec:vz} reviews the definition and boundaries of the VZ.
Section~\ref{sec:yield} presents our yield estimates under different
assumptions. Section~\ref{sec:disc} discusses the implications for
exoVenus science, synergies with Venus missions, and future
directions. Section~\ref{sec:con} provides concluding remarks.


\section{The PLATO Mission}
\label{sec:plato}

PLATO is ESA's third medium-class (M3) mission in the Cosmic Vision
2015--2025 programme \citep{rauer2014,rauer2025}, and will operate
from a halo orbit around the Sun-Earth Lagrange point~2 (L2).


\subsection{Instrument Design}
\label{sec:instrument}

The PLATO payload employs a modular multi-telescope architecture
consisting of 26 individual cameras: 24 ``normal'' cameras (NCAMs)
optimized for the 450--1000~nm bandpass and two ``fast'' cameras
(FCAMs) operating at higher cadence. Each NCAM is a refracting
telescope with a 120~mm entrance pupil diameter and a focal plane
equipped with four CCDs of $4510 \times 4510$ pixels, yielding an
individual field of view (FoV) of $\sim$1037~deg$^2$
\citep{ragazzoni2016a,nascimbeni2022}. The 24~NCAMs are arranged in
four groups of six co-pointing cameras, with the line of sight of each
group offset by 9.2$^\circ$ from the satellite boresight. This
configuration produces a total FoV of $\sim$2232~deg$^2$, covered by a
variable number of overlapping cameras: $\sim$301~deg$^2$ observed by
all 24~NCAMs, $\sim$247~deg$^2$ by 18, $\sim$735~deg$^2$ by 12, and
$\sim$949~deg$^2$ by 6 \citep{pertenais2021b,nascimbeni2025}. The
NCAMs acquire images at a 25-second cadence, while the FCAMs operate
at 2.5~seconds to monitor bright stars ($V = 4$--8).


\subsection{Stellar Sample and Target Fields}
\label{sec:fields}

PLATO will monitor at least 245,000~FGK dwarf and subgiant stars
brighter than $V = 13$, of which $\geq$15,000 high-priority targets
(the ``P1'' sample) have $V \leq 11$ and will be observed with
sufficient photometric precision for asteroseismic analysis
\citep{montalto2021}. An additional $\sim$5000 M~dwarfs with
$V < 16$ are included in the target sample.

The southern field LOPS2, centered at $(\alpha, \delta) = (06^{\rm
  h}21^{\rm m}14.5^{\rm s}, -47^\circ53'13'')$ (J2000), has been
selected as the first long-duration observation phase (LOP) field
\citep{nascimbeni2022,nascimbeni2025}. The LOPS2 field contains more
than 9000~FGK dwarfs and subgiants with $V < 11$ meeting photometric
noise requirements ($< 50$~ppm~hr$^{-1}$ in the 24-camera region), and
more than 159,000 such stars with $V < 13$.  The baseline observing
strategy consists of two LOP fields observed for two years each (the
``2+2'' scenario), though an extended 3-year pointing followed by a
1-year step-and-stare phase (``3+1'' scenario) is also under
consideration \citep{rauer2025}.


\subsection{Expected Planet Yield}
\label{sec:overview}

Using the Planet Yield for PLATO Estimator
\citep[PYPE;][]{matuszewski2023}, the expected total planet detection
yield for all sizes and orbital periods around stars brighter than $V
= 13$ is $\sim$4600, comparable to the total Kepler yield but around
substantially brighter host stars. For the subset of Earth-size
planets (0.8--1.25~$R_\oplus$), \citet{matuszewski2023} estimated a
minimum yield of $\sim$500 planets, approximately a dozen of which
would reside in the HZ of G-type stars with orbital periods in the
250--500~day range. In the bright P1 sample ($V \leq 11$), the yield
of small planets ($R_p < 2$~$R_\oplus$) across all orbital periods is
$\sim$770 \citep{rauer2025}. These estimates are sensitive to the
assumed planet occurrence rates, which represent the dominant source
of uncertainty in yield calculations
\citep{heller2022b,matuszewski2023}.


\section{The Venus Zone}
\label{sec:vz}


\subsection{Definition and Boundaries}
\label{sec:vzdef}

The VZ was formally defined by \citet{kane2014e} as the region around
a star where a terrestrial planet with an initially Earth-like
atmosphere would be expected to undergo a runaway greenhouse
transition, resulting in Venus-like surface conditions. The outer
boundary of the VZ corresponds to the runaway greenhouse limit as
computed by one-dimensional climate models
\citep{kasting1993a,kopparapu2013a,kopparapu2014}. For a solar-type
star ($T_{\rm eff} = 5780$~K), this limit occurs at an effective
stellar flux of $S_{\rm eff} = 1.107$~$S_\oplus$
\citep{kopparapu2014}, coinciding with the inner edge of the
conservative HZ. The inner boundary of the VZ is set by the stellar
flux level at which planetary atmospheres may become significantly
eroded, based on empirical scaling from atmospheric retention within
the Solar System \citep{zahnle2017}. The boundaries are parameterized
as functions of the host star effective temperature, making the VZ
applicable to stars across the main sequence.

\begin{figure*}
  \includegraphics[angle=270,width=\linewidth]{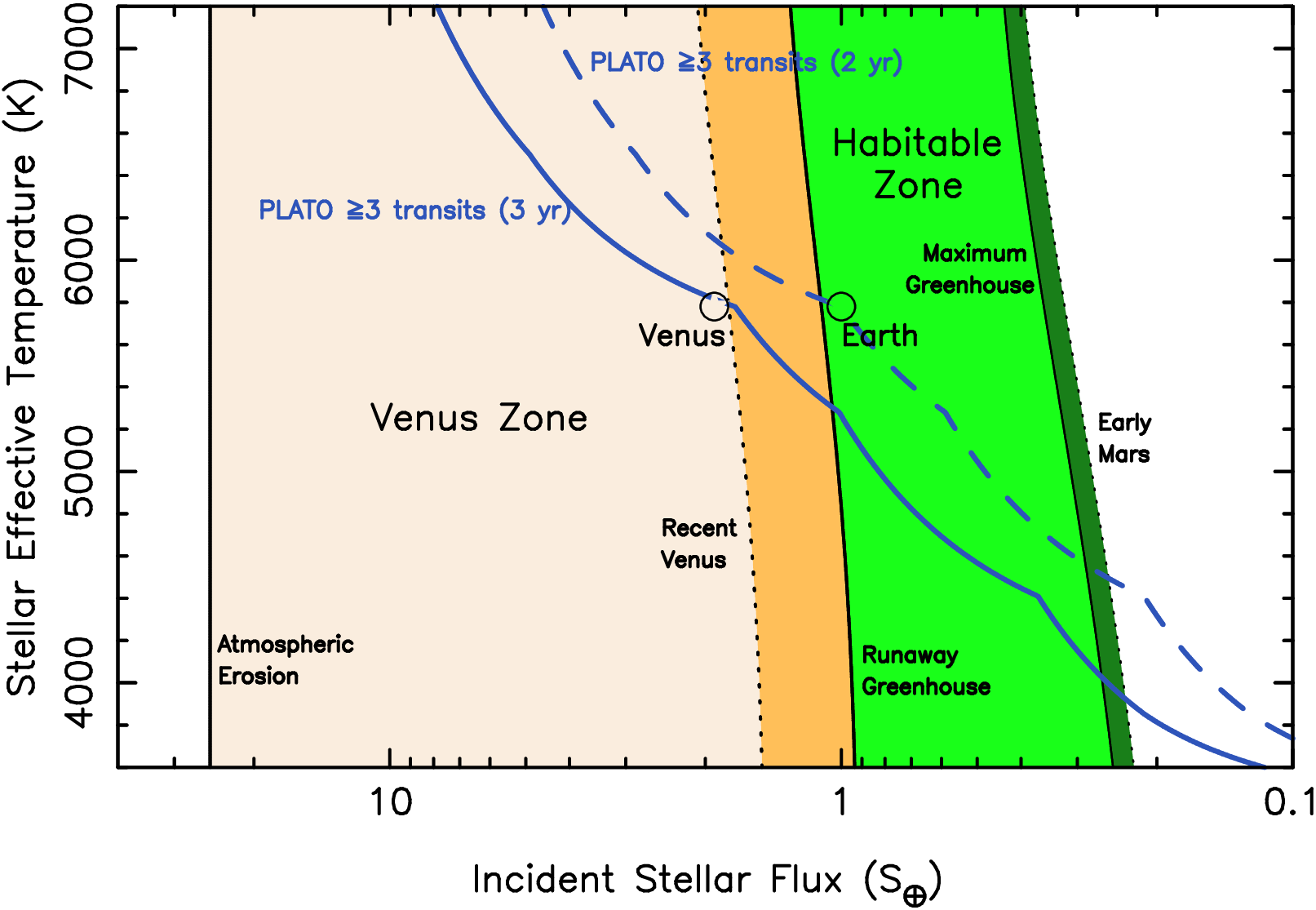}
  \caption{The Venus Zone (VZ) and Habitable Zone (HZ) shown in
    incident stellar flux ($S_{\rm eff}$) versus stellar effective
    temperature ($T_{\rm eff}$) space. The light orange region
    indicates the definite VZ and the darker orange region shows the
    optimistic VZ extension between the Recent Venus and runaway
    greenhouse limits. The bright and dark green regions show the
    conservative and optimistic HZ, respectively
    \citep{kopparapu2013a,kopparapu2014}. The solid and dashed blue
    curves show the minimum stellar flux detectable by PLATO with
    $\geq$3 transits in 2-year and 3-year continuous baselines,
    respectively; all planets to the left of these curves are
    accessible for a given stellar type. The locations of Venus and
    Earth in the solar system are marked. For K~dwarfs (lower portion
    of the plot), the entire VZ lies to the left of the PLATO 2-year
    detection horizon, while for G and F~dwarfs the outer VZ near the
    runaway greenhouse boundary requires a 3-year baseline for
    detection.}
  \label{fig:vz}
\end{figure*}

Between these limits, a terrestrial planet is expected to retain a
substantial atmosphere while receiving sufficient stellar irradiation
to trigger water loss via photolysis and hydrogen escape, followed by
the buildup of a dense CO$_2$ atmosphere, which is the evolutionary
pathway hypothesized for Venus
\citep{ingersoll1969c,kasting1988c,goldblatt2013,way2020}. The VZ thus
represents a physically motivated criterion for identifying potential
Venus analogs, analogous to the HZ for potential Earth analogs.
Figure~\ref{fig:vz} shows the VZ and HZ boundaries in the
$S_{\rm eff}$--$T_{\rm eff}$ plane, along with the PLATO detection
horizon discussed in Section~\ref{sec:yield}.

\subsection{Occurrence Rates}
\label{sec:occurrence}

Using Kepler data, \citet{kane2014e} identified 43 potential Venus
analogs with $R_p \leq 2~R_\oplus$ and derived occurrence rates
($\eta_{\rm Venus}$) of $0.32^{+0.05}_{-0.07}$ for M~dwarfs and
$0.45^{+0.06}_{-0.09}$ for GK~dwarfs, based on a sample of 43 VZ
planet candidates. While modest in size, this sample spans the F, G,
K, and M spectral types and the quoted uncertainties reflect the
corresponding Poisson and completeness errors. These values are
notably high, reflecting the geometric transit probability advantage
of shorter-period orbits compared to HZ planets. For comparison,
estimates of $\eta_\oplus$ for GK~dwarfs from Kepler data range from
$\sim$0.02 to $\sim$0.22, depending on the adopted HZ boundaries and
completeness corrections \citep{petigura2013a,bryson2021}. The higher
$\eta_{\rm Venus}$ values suggest that terrestrial planets in the VZ
may be more common than those in the HZ, a result with profound
implications for the overall frequency of habitable versus
uninhabitable terrestrial worlds.

\begin{figure*}
  \includegraphics[angle=270,width=\linewidth]{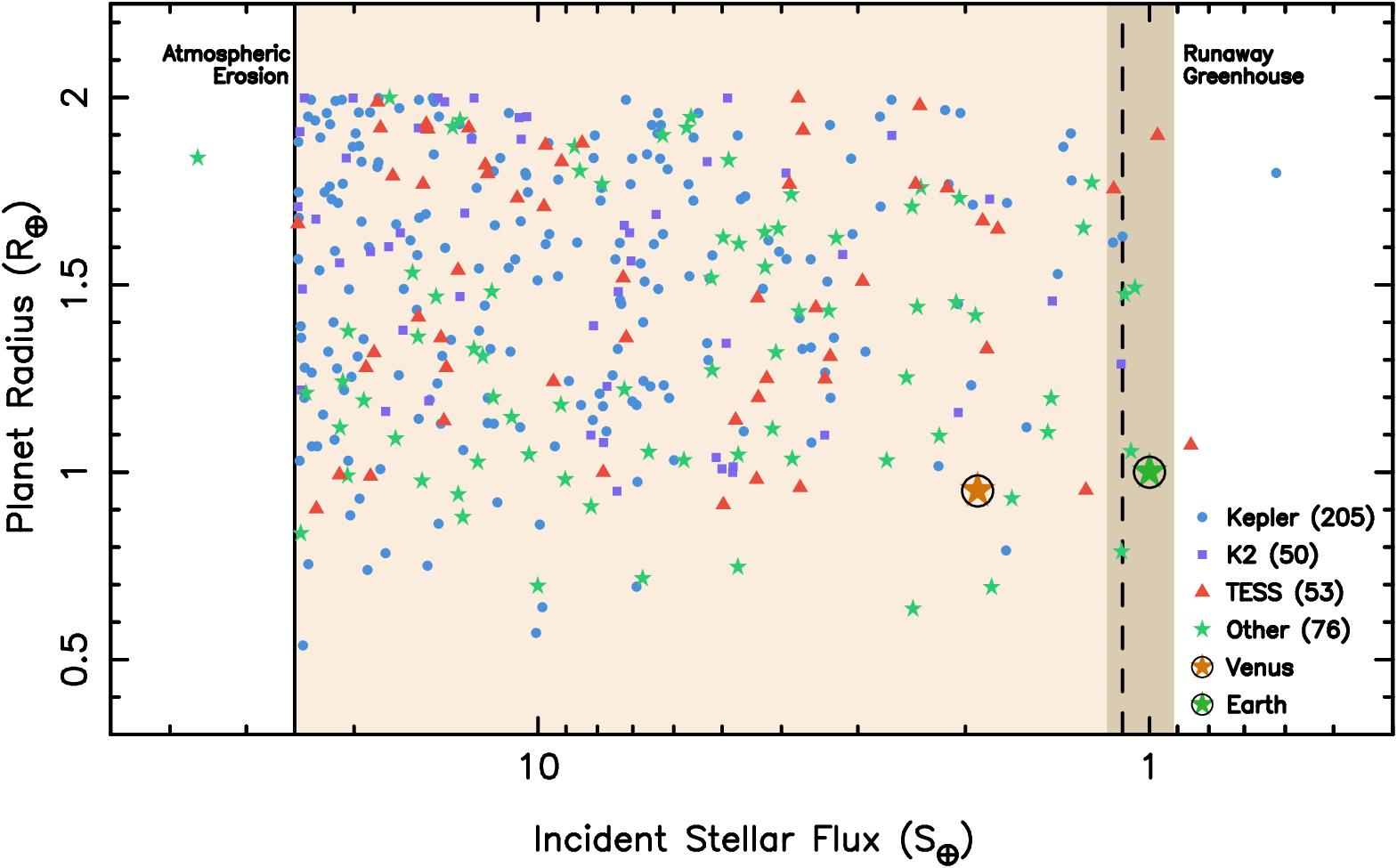}
  \caption{The current census of 384 VZ terrestrial planets in
    incident stellar flux ($S_{\rm eff}$) versus planet radius
    space. Points are colored by discovery mission: Kepler (blue
    circles), K2 (purple squares), TESS (red triangles), and other
    surveys including radial velocity (green diamonds). The light
    orange shading indicates the VZ. The darker band at the right edge
    of the VZ shows the range of the outer VZ boundary across the host
    star effective temperatures in the sample and the dashed line
    marks the solar value ($S_{\rm eff} = 1.107$~$S_\oplus$). The
    locations of Venus and Earth are marked as filled stars. The outer VZ near
    the runaway greenhouse boundary is sparsely populated at
    Earth-size radii, highlighting the region where PLATO will
    contribute the most new detections.}
  \label{fig:flux}
\end{figure*}

\citet{ostberg2023a} further characterized the VZ planet population,
finding that the period distribution of VZ terrestrial planets peaks
at $\sim$3--10~days, well within the detection sensitivity of transit
surveys. The median VZ planet radius was found to be
$\sim$1.3~$R_\oplus$, straddling the radius gap
\citep{fulton2017,vaneylen2018b}, which suggests a potential
connection between photoevaporative mass loss processes and the
conditions that produce Venus-like atmospheres. We extracted data from
the NASA Exoplanet Archive \citep{akeson2013,christiansen2025} on 2026
April 21, updating the total inventory of VZ planets with $R_p \leq
2$~$R_\oplus$ to 384 from previous estimates
\citep{ostberg2023a,kane2026a}. Where direct mass or radius
measurements are unavailable, estimates are derived from the
mass-radius relation of \citet{chen2017}. Figure~\ref{fig:flux} shows
our newly revised census of VZ terrestrial planets in incident
flux-radius space, colored by discovery mission. The census includes
planets on eccentric orbits that pass through the VZ. The population
is dominated by super-Earth radii ($R_p > 1.25$~$R_\oplus$) across all
flux levels, while Earth-size planets ($R_p \leq 1.25$~$R_\oplus$) are
comparatively sparse, particularly near the outer VZ boundary at the
runaway greenhouse limit.

\section{Venus Zone Yield Estimates for PLATO}
\label{sec:yield}


\subsection{Methodology}
\label{sec:method}

We estimate the PLATO VZ yield using two complementary approaches.
First, we apply an approximate VZ fraction to the published total
planet yield estimates from \citet{matuszewski2023}, scaling the
reported yield totals by the proportion of detectable orbital periods
that fall within the VZ (Section~\ref{sec:pype}). Second, we compute
the VZ yield independently using the empirical $\eta_{\rm Venus}$
occurrence rates from \citet{kane2014e}, the geometric transit
probability, and a simplified detection efficiency model applied
directly to the PLATO stellar sample
(Section~\ref{sec:eta}). Neither approach employs a full end-to-end
simulation of the PLATO detection pipeline; rather, they provide
complementary order-of-magnitude estimates whose consistency lends
confidence to the resulting yield range.

The PLATO detection horizon shown in Figure~\ref{fig:vz} is computed
analytically: for each stellar effective temperature, we adopt
empirical main-sequence mass and luminosity estimates interpolated
from the calibrations of \citet{pecaut2013}, then calculate the
maximum orbital period for which $\geq$3 transits occur within a
2-year (or 3-year) continuous baseline, converting to incident stellar
flux via Kepler's third law and $S_{\rm eff} = L_\star / a^2$. This
represents a geometric detection limit rather than a full
signal-to-noise threshold. These calculations assume circular orbits;
eccentricity would modify the time-averaged incident flux and could
shift planets into or out of the VZ depending on orbital phase.

The VZ for a solar-type star spans orbital periods of approximately
$\sim$33--338~days, corresponding to the inner VZ boundary ($S_{\rm
  eff} \approx 25~S_\oplus$, $a \approx 0.20$~AU) and the outer VZ
boundary at the runaway greenhouse limit ($S_{\rm eff} \approx
1.11~S_\oplus$, $a \approx 0.95$~AU). For K~dwarfs ($T_{\rm eff} \sim
4500$~K), the VZ period range shifts to $\sim$12--136~days, while for
F~dwarfs ($T_{\rm eff} \sim 6500$~K), it extends to
$\sim$72--696~days. These period ranges correspond to the VZ flux
boundaries shown in Figure~\ref{fig:vz}, where the PLATO detection
horizon (solid blue curve) indicates the minimum stellar flux
accessible with $\geq$3 transits in a 2-year baseline; planets to the
left of this curve are detectable. The expected
yield distribution across these period ranges is shown in
Figure~\ref{fig:period}.


\subsection{Yield from PYPE Scaling}
\label{sec:pype}

The PYPE calculations of \citet{matuszewski2023} provide published
estimates of the total PLATO planet yield as a function of planet
radius and orbital period. For the baseline 2+2~year observing
strategy, the total yield of terrestrial planets
($0.8$--$2.0~R_\oplus$) across all periods is approximately
1700--3000 planets around stars brighter than $V = 13$, depending on
the assumed occurrence rates (ranging from the empirical estimates of
\citealt{hsu2019c} to the theoretical predictions of the New
Generation Planetary Population Synthesis,
\citealt{emsenhuber2021a}).

\begin{figure*}
  \centering
  \includegraphics[angle=270,width=0.85\linewidth]{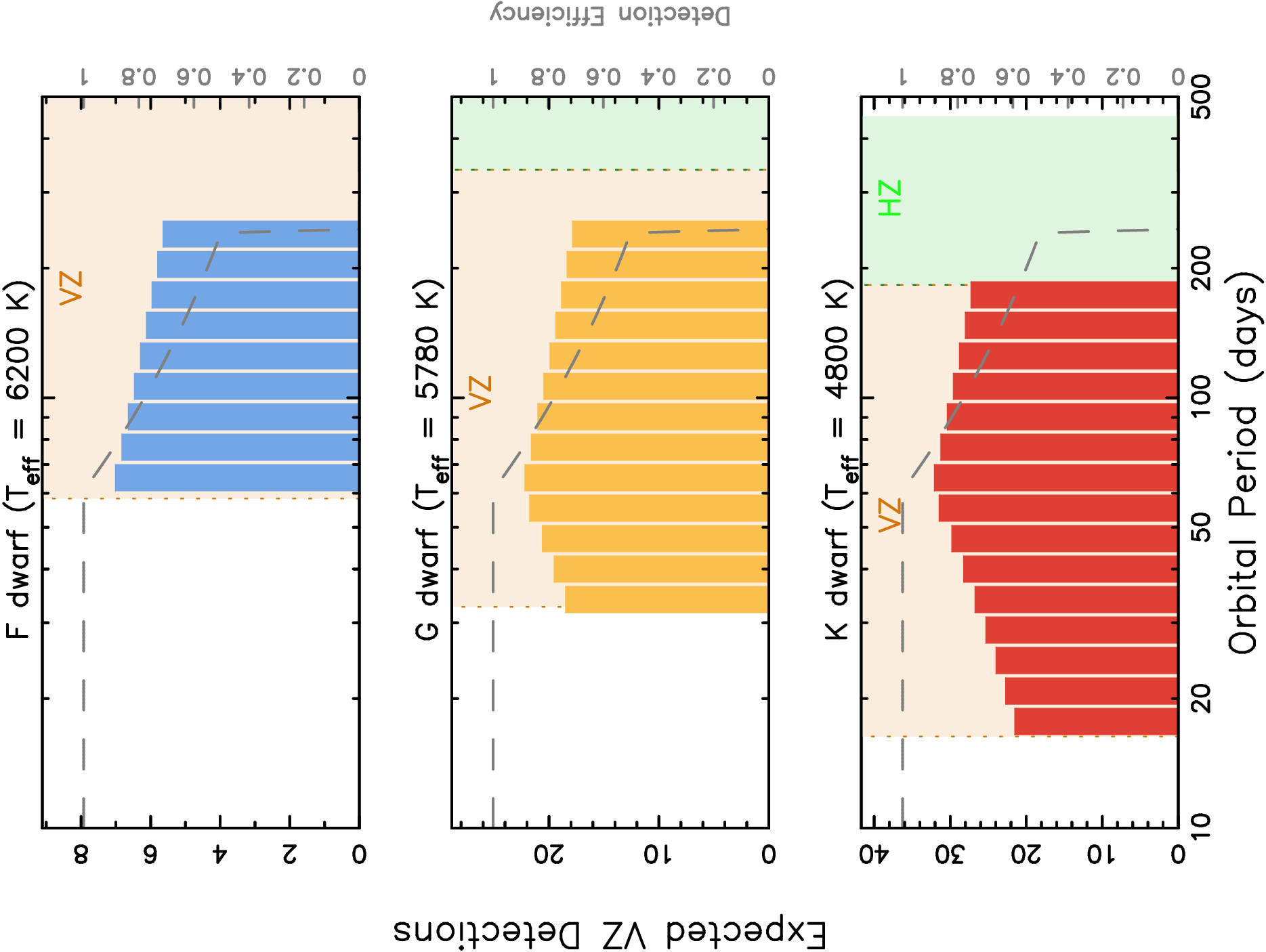}
  \caption{Expected VZ terrestrial planet yield as a function of
    orbital period for representative F~dwarf ($T_{\rm eff} =
    6200$~K; top), G~dwarf ($T_{\rm eff} = 5780$~K; middle), and
    K~dwarf ($T_{\rm eff} = 4800$~K; bottom) host stars. In each
    panel, the bars show the predicted number of VZ detections per
    period bin using $\eta_{\rm Venus} = 0.45$ \citep{kane2014e}, the
    geometric transit probability, and a detection efficiency model
    requiring $\geq$3 transits in a 2-year baseline (gray dashed
    curve; right axis). The light orange and light green shading
    indicate the VZ and HZ period ranges, respectively, with dotted
    vertical lines marking the boundaries. The VZ occupies
    progressively shorter periods for cooler stars, where the detection
    efficiency is higher. The K~dwarf VZ is fully accessible within a
    2-year baseline, while the outer VZ for G and F~dwarfs extends
    beyond the detection limit ($P_{\rm max} \approx 243$~days),
    requiring a 3-year pointing for complete coverage.}
  \label{fig:period}
\end{figure*}

We estimate the VZ fraction of this total yield by numerically
integrating the detection-weighted yield across all orbital periods
and computing the fraction that falls within the VZ. The
detection-weighted yield per unit log-period scales as
\begin{equation}
  \frac{dN}{d\log P} \propto \eta(P) \times p_{\rm tr}(P) \times \epsilon(P)
\end{equation}
where $\eta(P)$ is the occurrence rate, $p_{\rm tr} \propto P^{-2/3}$
is the geometric transit probability, and $\epsilon(P)$ is the
detection efficiency (the probability that PLATO detects a transiting
planet, given sufficient signal-to-noise and $\geq$3 observed transit
events). The VZ fraction depends sensitively on the occurrence rate
slope: if $\eta$ is flat in $\log P$, the strong transit probability
bias toward short periods concentrates most detections at $P <
30$~days, yielding $f_{\rm VZ} \sim 5\%$. However, Kepler data show
that terrestrial planet occurrence rates rise significantly with
period in the 1--300~day range, with $dN/d\log P \propto P^{0.5}$
\citep{hsu2019c}. This rising occurrence rate shifts the
detection-weighted yield toward longer periods, substantially
increasing the VZ fraction. For $dN/d\log P \propto P^{0.5}$, we
obtain $f_{\rm VZ} \approx 10$--$27\%$ across F, G, and K spectral
types, with an FGK-weighted mean of $\sim$21\%. We therefore adopt
$f_{\rm VZ} \approx 15$--25\% as a representative range, corresponding
to moderate-to-steep occurrence rate slopes. This range encompasses
power law indices from $\alpha \approx 0.3$ to 0.7, which brackets the
reported uncertainty on the \citet{hsu2019c} slope. Adopting a central
value of 20\% and applying this to the PYPE yield totals:
\begin{equation}
  N_{\rm VZ} = f_{\rm VZ} \times N_{\rm terr} \approx 0.20 \times
  (1700\text{--}3000) \approx 340\text{--}600
\end{equation}
where $N_{\rm terr}$ is the total terrestrial planet yield. However,
this estimate includes the full terrestrial size range up to
$2~R_\oplus$, which encompasses both true rocky planets and planets
with significant volatile envelopes that may not be genuine Venus
analogs. Restricting to Earth-size planets ($0.8$--$1.25~R_\oplus$),
which are the most relevant comparators to Venus ($R_{\rm Venus} =
0.95~R_\oplus$), reduces the yield by a factor of $\sim$3--5, giving:
\begin{equation}
N_{\rm VZ, Earth\text{-}size} \approx 70\text{--}200
\end{equation}
for all stars with $V < 13$. The resulting yield distribution as a
function of orbital period is shown in Figure~\ref{fig:period} for
representative F, G, and K dwarf host stars.

\subsection{Yield from Occurrence Rate Analysis}
\label{sec:eta}

As a complementary estimate, we compute the VZ yield using $\eta_{\rm
  Venus}$ applied directly to the PLATO stellar sample. This approach
does not rely on the PYPE yield totals and instead combines occurrence
rates, geometric transit probabilities, and a simplified detection
efficiency model. For the P1 sample of $\sim$15,000 FGK dwarfs with $V
\leq 11$, and adopting $\eta_{\rm Venus} = 0.45$ for GK dwarfs
\citep{kane2014e}, the expected number of VZ terrestrial planets is:
\begin{equation}
N_{\rm VZ, P1} = N_\star \times \eta_{\rm Venus} \times p_{\rm tr} \approx 15{,}000 \times 0.45 \times p_{\rm tr}
\end{equation}
where $p_{\rm tr}$ is the geometric transit probability averaged over
the VZ. For the VZ of a G2V star, $p_{\rm tr}$ ranges from $\sim$0.5\%
at the outer boundary ($a \approx 0.95$~AU) to $\sim$2\% at the inner
boundary ($a \approx 0.20$~AU), with a flux-weighted mean of
$\sim$1.5\%. This value represents an FGK-weighted average; the mean
$p_{\rm tr}$ varies from $\sim$1\% for F~dwarfs (whose VZ lies at
larger orbital distances) to $\sim$2.5\% for K~dwarfs, as reflected in
the spectral-type-resolved yields shown in
Figure~\ref{fig:period}. This yields:
\begin{equation}
N_{\rm VZ, P1} \approx 15{,}000 \times 0.45 \times 0.015 \approx 100
\end{equation}
We model the PLATO detection efficiency as $\epsilon = \min(1,\;
0.5\sqrt{n_{\rm tr}/3})$ for planets receiving $\geq$3 transits during
the observing baseline, and $\epsilon = 0$ otherwise, where $n_{\rm
  tr}$ is the number of transits. This approximation captures the
basic scaling of signal-to-noise with transit count but does not
incorporate the detailed PLATO multi-camera noise model. Varying the
prefactor in this prescription from 0.3 to 0.7 (bracketing the adopted
value of 0.5) changes the P1 yield by approximately a factor of two,
which is smaller than the factor-of-several uncertainty introduced by
the occurrence rate. Averaged over the VZ period distribution, this
efficiency prescription reduces the yield by a characteristic factor
of $\sim$0.5--0.85, giving a corrected estimate of $\sim$50--85 VZ
terrestrial planets in the P1 sample alone.

Extending to the full P5 sample ($V \leq 13$, $\sim$245,000 stars),
the larger number of targets partially compensates for the reduced
characterization potential:
\begin{equation}
N_{\rm VZ, P5} \approx 245{,}000 \times 0.45 \times 0.015 \times \epsilon_{\rm P5}
\end{equation}
where $\epsilon_{\rm P5}$ is the mean detection efficiency for the P5
sample. Applying the same efficiency model as for the P1 sample
($\epsilon \approx 0.5$--0.85 across the VZ period range), we obtain
$N_{\rm VZ, P5} \approx 825$--1400 VZ terrestrial planets. Since
$\eta_{\rm Venus}$ is already defined for planets with $R_p <
2~R_\oplus$, no additional size correction is required. For the
Earth-size subset, adopting the observed fraction of Earth-size
planets in the VZ ($\sim$26\%; \citealt{ostberg2023a}), we obtain
$\sim$215--365 Earth-size VZ planets in the P5 sample. These estimates
are upper bounds: the simplified detection efficiency model does not
incorporate brightness-dependent photometric noise, which will reduce
the sensitivity to small planets around the fainter P5 stars ($V =
11$--13) relative to the bright P1 sample.


\subsection{Summary of Yield Estimates}
\label{sec:summary}

Table~\ref{tab:yield} summarizes our VZ yield estimates under the
different approaches and assumptions described above. An important
caveat is that both approaches rely on planet occurrence rates derived
from the Kepler sample, which targeted stars at $V \sim 12$--16 in a
single field toward Cygnus-Lyra \citep{borucki2010a}. The PLATO P1
sample, by contrast, will observe brighter ($V \leq 11$), typically
nearer FGK dwarfs distributed across different Galactic sight
lines. If occurrence rates depend on stellar metallicity, age, or
Galactic environment, the Kepler-derived values may not transfer
directly. However, studies of small planet occurrence rates from
Kepler show a weak dependence on host star metallicity for planets
with $R_p < 2~R_\oplus$ \citep{buchhave2014}, and both samples are
dominated by thin-disk FGK dwarfs with broadly similar
properties. Furthermore, TESS has confirmed that small planet
occurrence rates around bright, nearby stars are consistent with
Kepler estimates \citep{kunimoto2020b}. We therefore expect the
Kepler-derived $\eta_{\rm Venus}$ to be applicable to the PLATO sample
at the level of precision of our yield estimates, though PLATO will
ultimately measure VZ occurrence rates for its own stellar sample
directly.

The $\eta_{\rm Venus}$ estimates for the P5 sample yield higher
numbers (825--1400 all terrestrial) than the PYPE-based scaling
(340--600), reflecting both the higher occurrence rates adopted in the
former and the fact that the simplified detection efficiency model
does not account for reduced sensitivity to small planets around
fainter stars. The $\eta_{\rm Venus}$ P5 estimates are therefore upper
bounds, and we adopt the conservative PYPE range as the most
defensible overall estimate. This conservative range combines the
lower PYPE terrestrial yield ($\sim$1700 planets, based on the
empirical rates of \citealt{hsu2019c}) with the lower end of the
$f_{\rm VZ}$ range ($\sim$10--16\%), giving $\sim$170--280 VZ
terrestrial planets ($0.8$--$2.0~R_\oplus$), including $\sim$40--80
Earth-size ($0.8$--$1.25~R_\oplus$) planets. For the bright P1 sample
($V \leq 11$), the $\eta_{\rm Venus}$-based estimate gives
$\sim$50--85 terrestrial and $\sim$13--22 Earth-size VZ
detections. Considering both size cuts and yield methods, the full
plausible range spans $\sim$40--280 VZ terrestrial planets, from
Earth-size conservative lower limits to all-terrestrial upper
limits. These yields are sensitive to the choice of observing
strategy, as explored in Figure~\ref{fig:strategy}. The predicted
brightness distribution is shown in Figure~\ref{fig:vmag}, which
contrasts the current VZ census (peaked at $V \sim 14$--15) with the
PLATO yield (peaked at $V \sim 11$--13).

\begin{deluxetable}{lcc}
\tablecaption{Estimated terrestrial VZ yield from PLATO.\label{tab:yield}}
\tablewidth{0pt}
\tablehead{
\colhead{Sample / Method} & \colhead{$N_{\rm VZ}$} & \colhead{$N_{\rm VZ}$} \\
\colhead{} & \colhead{($0.8$--$2.0~R_\oplus$)} & \colhead{($0.8$--$1.25~R_\oplus$)}
}
\startdata
P1 ($V \leq 11$), $\eta_{\rm Venus}$ & 50--85 & 13--22 \\
P5 ($V \leq 13$), $\eta_{\rm Venus}$ & 825--1400 & 215--365 \\
PYPE ($V \leq 13$), all terrestrial & 340--600 & 70--200 \\
PYPE ($V \leq 13$), conservative & 170--280 & 40--80 \\
\enddata
\tablecomments{The ``conservative'' PYPE estimate adopts the lower
  occurrence rates from \citet{hsu2019c}. The ``all terrestrial''
  estimate uses the NGPPS rates of \citet{emsenhuber2021a}.}
\end{deluxetable}


\section{Discussion}
\label{sec:disc}


\subsection{Implications for the Earth-Venus Divergence}
\label{sec:divergence}

\begin{figure*}
  \includegraphics[angle=270,width=\linewidth]{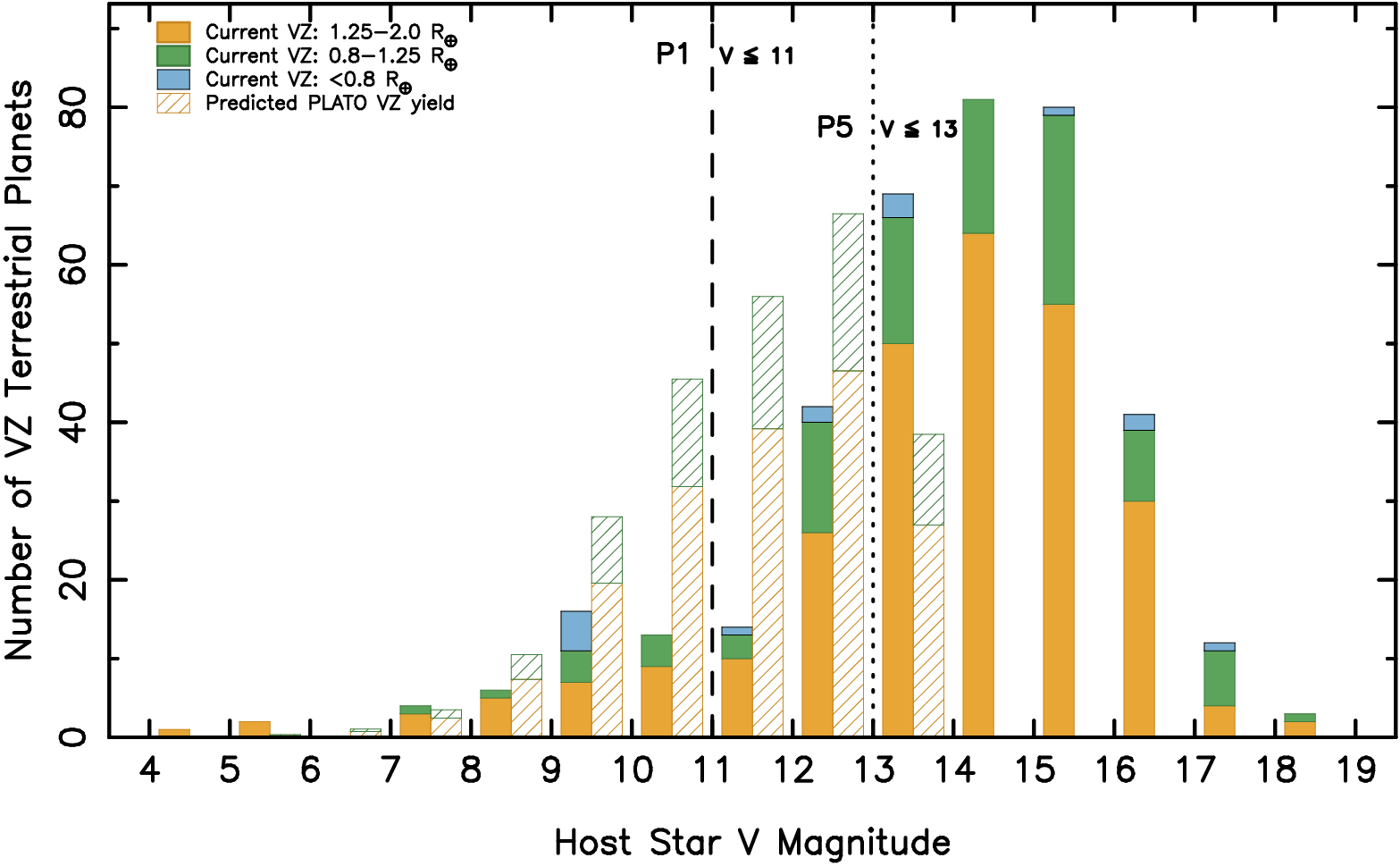}
  \caption{Distribution of VZ terrestrial planets as a function of
    host star $V$ magnitude. In each magnitude bin, the left bars show
    our newly revised census of 384 VZ planets, stacked by radius:
    super-Earth ($1.25$--$2.0~R_\oplus$; orange), Earth-size
    ($0.8$--$1.25~R_\oplus$; green), and sub-Earth ($< 0.8~R_\oplus$;
    blue). The right bars (hatched) show the predicted PLATO VZ yield,
    normalized to $\sim$250 total detections, consistent with the
    conservative PYPE estimate (Table~\ref{tab:yield}), with the same radius
    color scheme. The dashed and dotted vertical lines indicate the
    PLATO P1 ($V \leq 11$) and P5 ($V \leq 13$) sample magnitude
    limits, respectively. The current census is dominated by faint
    Kepler targets ($V \sim 14$--16), while PLATO will preferentially
    detect VZ planets around brighter stars accessible for radial
    velocity follow-up and atmospheric characterization.}
  \label{fig:vmag}
\end{figure*}

The detection of $\sim$170--280 VZ terrestrial planets by PLATO
($\sim$40--80 at Earth-size radii) would
represent a transformative advance in our ability to study the
Earth-Venus divergence through a statistical lens. Currently, only
$\sim$384 confirmed terrestrial VZ planets are known (see
Section~\ref{sec:occurrence}), and the majority orbit faint stars ($V
> 13$) that are poorly suited for atmospheric characterization
(Figure~\ref{fig:vmag}). The PLATO sample will provide VZ planets
around significantly brighter host stars, enabling mass measurements
via radial velocity follow-up to precisions of $\leq$10\%
\citep{rauer2025} and transmission spectroscopy observations with JWST
and future facilities such as the Extremely Large Telescope (ELT), the
Large Interferometer For Exoplanets (LIFE; \citealt{quanz2022a}), and
HWO.

A key scientific question is whether the transition from habitable to
Venus-like conditions occurs sharply at the runaway greenhouse
boundary or exhibits a more gradual dependence on stellar flux, planet
mass, atmospheric composition, and other factors
\citep{kane2014e,kane2019d,schlecker2024}. As shown in
Figure~\ref{fig:flux}, the outer VZ near the runaway greenhouse limit
is currently sparsely populated at Earth-size radii, limiting our
ability to probe this transition observationally. The PLATO VZ sample,
combined with atmospheric constraints, will allow us to test whether
all planets above the runaway greenhouse flux threshold exhibit
Venus-like atmospheres, or whether a significant fraction retain
temperate conditions, potentially extending the HZ inward
\citep{leconte2013c,way2020,turbet2021}. A key unknown is what
fraction of VZ terrestrial planets retain substantial atmospheres at
all: recent JWST observations suggest that some highly irradiated
rocky planets may lack atmospheres entirely \citep{greene2023}, while
theoretical models predict that planets in the moderate-flux VZ regime
should retain dense envelopes. PLATO's bright-star VZ detections will
provide the sample needed to measure this retention fraction
observationally.


\subsection{The Role of Stellar Type}
\label{sec:stellar}

The PLATO target sample is dominated by FGK dwarfs, which provide the
most direct comparison to the Sun-Venus system. For G-type stars, the
VZ orbital periods ($\sim$33--338~days) are partially matched to
PLATO's detection sensitivity over a 2-year baseline, which can detect
planets with $\geq$3 transits out to $\sim$243~days. The inner and
middle VZ are fully accessible, but the critical outer VZ near the
runaway greenhouse boundary ($P \sim 243$--338~days) requires a 3-year
pointing to ensure $\geq$3 transits. For K~dwarfs, the VZ shifts to
shorter periods ($\sim$12--136~days) where detection efficiency is
even higher, and a 2-year baseline captures the full VZ, potentially
yielding a disproportionately large number of VZ detections relative
to HZ detections. For F~dwarfs, the VZ extends to longer periods
($\sim$72--696~days) where detection becomes significantly more
challenging, though the inner VZ remains accessible. This spectral
type dependence is evident in Figure~\ref{fig:vz}, where the PLATO
2-year detection curve lies entirely to the right of the VZ for
K~dwarfs but cuts through the outer VZ for G and F~dwarfs. The corresponding yield
distributions in Figure~\ref{fig:period} further illustrate this
effect: the K~dwarf panel shows detections spanning the full VZ with
uniformly high detection efficiency, while the F~dwarf panel shows
yields truncated at long periods where the efficiency drops
precipitously.

The stellar type distribution of VZ detections will provide insight
into how the runaway greenhouse boundary depends on spectral energy
distribution, particularly the UV flux that drives photolysis of water
vapor in the upper atmosphere \citep{wordsworth2013b,luger2015b}.
PLATO's asteroseismic capabilities will further enable precise stellar
age determinations, allowing the VZ population to be studied as a
function of system age for the first time. For planets in the outer VZ
that receive only two transits during the 2-year baseline, radial
velocity campaigns can confirm the orbital period and measure the
planetary mass, provided the host star is sufficiently bright and
chromospherically quiet. Such two-transit candidates will constitute
valuable targets for extended ground-based follow-up.

\begin{figure*}
  \includegraphics[angle=270,width=\linewidth]{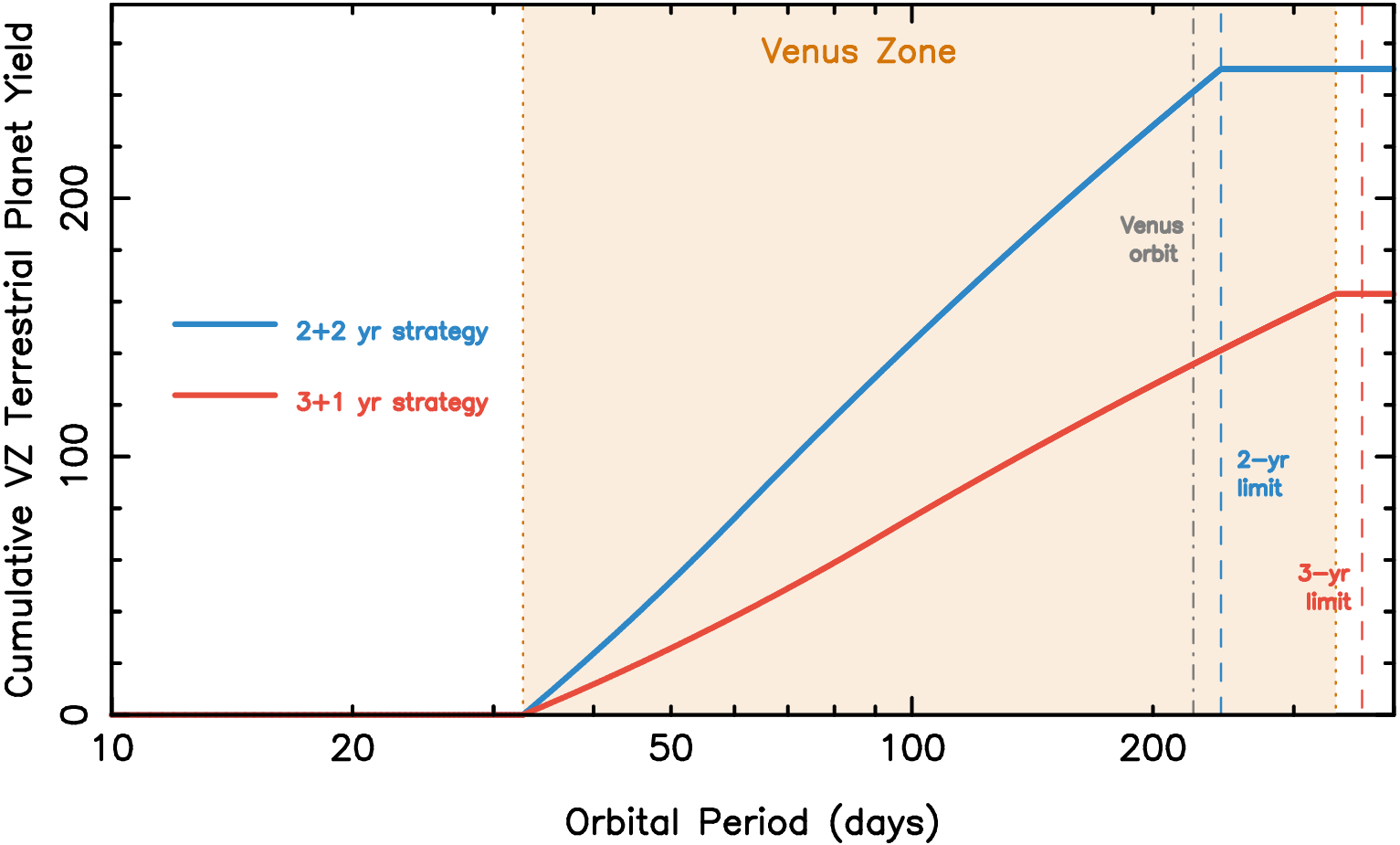}
  \caption{Cumulative VZ terrestrial planet yield as a function of
    orbital period for the 2+2~yr (blue) and 3+1~yr (red) observing
    strategies, computed for a G2V host star. The light orange shading
    indicates the VZ, with dotted lines marking its boundaries. The
    dashed blue and red lines show the maximum orbital period
    detectable with $\geq$3 transits in 2-year ($P_{\rm max} \approx
    243$~days) and 3-year ($P_{\rm max} \approx 365$~days) baselines,
    respectively. The dash-dot gray line marks the orbital period of
    Venus (224.7~days). The 2+2 strategy produces a higher total yield
    because it observes two independent fields, but truncates at
    $\sim$243~days. The 3+1 strategy extends sensitivity to the outer
    VZ near the runaway greenhouse boundary, though the step-and-stare
    phase contributes negligibly to the VZ yield for G~dwarfs because
    its 60-day baselines fall shortward of the VZ inner boundary.}
  \label{fig:strategy}
\end{figure*}

\subsection{Synergies with Venus Missions}
\label{sec:synergies}

The PLATO mission timeline is fortuitously aligned with a resurgence
of Venus exploration. NASA's DAVINCI mission \citep{garvin2022} will
perform in-situ measurements of Venus's atmospheric composition during
descent, while VERITAS \citep{cascioli2021} will provide global radar
imaging and gravity mapping. ESA's EnVision mission
\citep{widemann2023} will conduct comprehensive orbital observations
of Venus's surface, atmosphere, and interior. All three of these
missions have planned deployments in the early to mid 2030s. In
addition, India's planned Venus Orbiter Mission (VOM), also known as
Shukrayaan-1, will conduct radar mapping and IR reconnaissance in
2028--2031 from an elliptical orbit \citep{limaye2023}. The ensemble
of orbital measurements at Venus from VOM, EnVision, and VERITAS will
paint a comprehensive picture of the surface and interior using radar
and gravity measurements and link them to global properties of the
atmosphere, mostly from the cloud deck up to the margins of
space. DAVINCI's in situ campaign will produce global chemical context
from the clouds to the surface with detailed measurements every few
hundred meters together with high accuracy pressure and temperature
(at 0.1~K resolution) all the way to the surface over a mountain
region known as Alpha Regio. Whether the deep atmosphere of Venus is
in an equilibrium state or dynamically reactive with the surface
producing signatures within trace gases involving SO$_2$, CO, Fe, and
others is unknown but suspected, and the directly measured partial
pressure of oxygen (fO$_2$) that DAVINCI will provide will further
anchor the interpretations of Venus atmospheric evolution to its
present state. Thus, these missions will provide ground-truth data on
Venus's current state that can be used to calibrate and validate
atmospheric retrieval models applied to exoVenus candidates detected
by PLATO \citep{horner2020b,kane2021d,kane2024b}.

The synergy operates in both directions: Venus mission data will
inform the interpretation of exoVenus spectra, while the statistical
properties of the exoVenus population will provide context for
understanding whether Venus's evolutionary pathway is typical or
exceptional among terrestrial planets in the VZ
\citep{kane2019d,kane2024b}. Having sub-cloud details of the Venusian
atmospheric chemistry, fO$_2$, lapse rate ($dT/dz$), and whether the
nearest surface trace gas chemistry reflects a more reactive chemical
environment or an equilibrium situation will influence scientific
opportunities for the VZ. This bidirectional approach, using Venus as
a spectral ground truth for exoplanet characterization and using
exoplanets to place Venus in a statistical context, represents a
powerful paradigm for advancing our understanding of terrestrial
planet habitability. Private ventures such as the Rocket Lab mission
to Venus \citep{seager2022a}, which aims to conduct an atmospheric
entry probe in the late 2020s, may provide additional early data on
the Venusian atmosphere that could further inform exoVenus
interpretation.


\subsection{Observing Strategy Considerations}
\label{sec:strategy}

The choice between the 2+2 and 3+1 observing strategies has
significant implications for the VZ yield. The 2+2 strategy observes
two fields for two years each, maximizing the total number of
monitored stars and therefore the overall planet yield at periods
accessible within a 2-year baseline. The 3+1 strategy dedicates three
years to a single field, extending sensitivity to longer-period planets
including those in the outer VZ and HZ of G-type stars, followed by
one year of shorter step-and-stare observations that sample many
additional fields at reduced period sensitivity
\citep{rauer2025,matuszewski2023}.

For VZ planets specifically, the 3+1 strategy offers advantages at
the outer boundary (periods of $\sim$243--338~days for G stars), where
the additional year of continuous monitoring significantly improves the
probability of detecting three or more transits. This outer VZ region
is inaccessible to a 2-year pointing ($P_{\rm max} \approx 243$~days
for $\geq$3 transits) but is fully covered by a 3-year baseline
($P_{\rm max} \approx 365$~days). However, the step-and-stare phase
in the 3+1 scenario contributes negligibly to VZ detections around
solar-type stars: the 60-day step-and-stare baselines limit detection to $P <
20$~days, which falls shortward of the VZ inner boundary for G~dwarfs
($P \sim 33$~days). The step-and-stare phase may contribute short-period VZ
detections only around late K and M~dwarfs, where the VZ extends to
sufficiently short periods. Figure~\ref{fig:strategy} shows the
cumulative VZ yield as a function of orbital period for both
strategies, illustrating the higher total yield of the 2+2 scenario
(which surveys twice the sky area) and the extended period coverage of
the 3+1 scenario near the runaway greenhouse boundary.


\subsection{Comparison with TESS and Kepler}
\label{sec:comparison}

The existing census of VZ planets from Kepler and TESS provides a
baseline against which PLATO's expected contribution can be
assessed. Kepler detected the majority of currently known VZ
candidates, but most orbit faint stars ($V > 14$) that are difficult
to characterize further. TESS has identified VZ planets around
bright nearby stars, but its short observing baselines (27~days per
sector in the primary mission) limit its sensitivity to the outer VZ.
Figure~\ref{fig:vmag} illustrates this brightness distribution: the
current VZ census peaks at $V \sim 14$--15, dominated by Kepler
detections, while the predicted PLATO yield peaks at $V \sim 11$--13,
shifting the population decisively toward characterizable host stars.

PLATO occupies a unique parameter space in that it combines
Kepler-like continuous monitoring durations with TESS-like brightness
sensitivity ($V \leq 11$ for the P1 sample). This combination is
critical for VZ science because it enables the detection of VZ planets
at all stellar flux levels within the zone: from the hottest,
shortest-period planets near the atmospheric erosion boundary to the
critical outer boundary where the transition from habitable to
Venus-like conditions may occur. The latter region is of particular
scientific interest because it probes the conditions under which
runaway greenhouse transitions are initiated, and it requires the long
monitoring baselines that only PLATO (among current or planned
missions) can provide for a large sample of bright stars.


\section{Conclusions}
\label{sec:con}

Although there is likely a continuum of planetary environments that
exist within the exoplanet population, those cases that may be
ground-truthed through a direct comparison to Solar System bodies are
especially useful. Venus is particularly valuable in this regard given
its dramatic past divergence with Earth, the study of which will be
greatly informed through an expansion of the exoVenus population
around bright stars. We have estimated the expected yield of Venus
Zone terrestrial planets from ESA's PLATO mission, finding that the
mission should detect $\sim$170--280 VZ terrestrial planets
($0.8$--$2.0~R_\oplus$), including $\sim$40--80 at Earth-size radii,
under conservative assumptions. For the bright P1 sample ($V \leq
11$), we estimate $\sim$50--85 terrestrial and $\sim$13--22 Earth-size
VZ detections, enabling radial velocity mass determination,
asteroseismic age dating (for the F and G~dwarf subset where
oscillations are most readily detected), and atmospheric
characterization of favorable targets with current and future
facilities. PLATO will increase the number of bright, characterizable
VZ terrestrial planets by an order of magnitude compared to current
inventories, providing the statistical power needed to constrain the
frequency and conditions of runaway greenhouse outcomes. The VZ yield
is substantial even under conservative occurrence rate assumptions,
because the VZ spans a period range where PLATO's detection efficiency
is high and the geometric transit probability is significantly larger
than for HZ planets. Furthermore, the combination of PLATO's bright
stellar sample with asteroseismic age determinations will enable the
first systematic study of VZ planet properties as a function of system
age, probing the timescales of atmospheric evolution and climate
divergence. The PLATO VZ sample will also be highly synergistic with
data from the DAVINCI, VERITAS, and EnVision missions to Venus,
enabling a bidirectional calibration between solar system ground truth
and exoplanet statistics.

Future work should refine these yield estimates by integrating the
detailed period- and radius-dependent detection efficiency tables from
the PYPE pipeline \citep{matuszewski2023} with the VZ boundaries
computed for individual target stars, replacing the approximate
scaling approach adopted here. Such an analysis would account for the
PLATO multi-camera noise model, the specific stellar populations in
the LOPS2 and LOPN1 fields, and the period-dependent completeness
corrections that our simplified detection efficiency model does not
capture. However, the dominant uncertainty in VZ yield estimates is
the occurrence rate itself, not the detection efficiency, and the
consistency between our two complementary approaches
(Sections~\ref{sec:pype} and \ref{sec:eta}) suggests that the
conservative expectation of tens to hundreds of PLATO VZ detections is
robust to the level of approximation employed. The
PLATO mission will provide an unprecedented opportunity to study the
boundary conditions for planetary habitability, not only by searching
for Earth analogs in the HZ, but by characterizing the far more
numerous population of potential Venus analogs that populate the inner
regions of planetary systems. Understanding the prevalence and
diversity of Venus-like worlds is essential for placing Earth's
climate evolution in a broader context and for assessing whether our
planet's habitability is the rule or the exception.


\section*{Acknowledgements}

This research has made use of the NASA Exoplanet Archive, which is
operated by the California Institute of Technology, under contract
with the National Aeronautics and Space Administration under the
Exoplanet Exploration Program. The results reported herein benefited
from collaborations and/or information exchange within NASA's Nexus
for Exoplanet System Science (NExSS) research coordination network
sponsored by NASA's Science Mission Directorate.



\begin{thebibliography}{}
\expandafter\ifx\csname natexlab\endcsname\relax\def\natexlab#1{#1}\fi
\providecommand{\url}[1]{\href{#1}{#1}}
\providecommand{\dodoi}[1]{doi:~\href{http://doi.org/#1}{\nolinkurl{#1}}}
\providecommand{\doeprint}[1]{\href{http://ascl.net/#1}{\nolinkurl{http://ascl.net/#1}}}
\providecommand{\doarXiv}[1]{\href{https://arxiv.org/abs/#1}{\nolinkurl{https://arxiv.org/abs/#1}}}

\bibitem[{{Akeson} {et~al.}(2013){Akeson}, {Chen}, {Ciardi}, {Crane}, {Good},
  {Harbut}, {Jackson}, {Kane}, {Laity}, {Leifer}, {Lynn}, {McElroy}, {Papin},
  {Plavchan}, {Ram{\'\i}rez}, {Rey}, {von Braun}, {Wittman}, {Abajian}, {Ali},
  {Beichman}, {Beekley}, {Berriman}, {Berukoff}, {Bryden}, {Chan}, {Groom},
  {Lau}, {Payne}, {Regelson}, {Saucedo}, {Schmitz}, {Stauffer}, {Wyatt}, \&
  {Zhang}}]{akeson2013}
{Akeson}, R.~L., {Chen}, X., {Ciardi}, D., {et~al.} 2013, \pasp, 125, 989,
  \dodoi{10.1086/672273}

\bibitem[{{Barstow} {et~al.}(2016){Barstow}, {Aigrain}, {Irwin}, {Kendrew}, \&
  {Fletcher}}]{barstow2016a}
{Barstow}, J.~K., {Aigrain}, S., {Irwin}, P.~G.~J., {Kendrew}, S., \&
  {Fletcher}, L.~N. 2016, \mnras, 458, 2657, \dodoi{10.1093/mnras/stw489}

\bibitem[{{Borucki} {et~al.}(2010){Borucki}, {Koch}, {Basri}, {Batalha},
  {Brown}, {Caldwell}, {Caldwell}, {Christensen-Dalsgaard}, {Cochran},
  {DeVore}, {Dunham}, {Dupree}, {Gautier}, {Geary}, {Gilliland}, {Gould},
  {Howell}, {Jenkins}, {Kondo}, {Latham}, {Marcy}, {Meibom}, {Kjeldsen},
  {Lissauer}, {Monet}, {Morrison}, {Sasselov}, {Tarter}, {Boss}, {Brownlee},
  {Owen}, {Buzasi}, {Charbonneau}, {Doyle}, {Fortney}, {Ford}, {Holman},
  {Seager}, {Steffen}, {Welsh}, {Rowe}, {Anderson}, {Buchhave}, {Ciardi},
  {Walkowicz}, {Sherry}, {Horch}, {Isaacson}, {Everett}, {Fischer}, {Torres},
  {Johnson}, {Endl}, {MacQueen}, {Bryson}, {Dotson}, {Haas}, {Kolodziejczak},
  {Van Cleve}, {Chandrasekaran}, {Twicken}, {Quintana}, {Clarke}, {Allen},
  {Li}, {Wu}, {Tenenbaum}, {Verner}, {Bruhweiler}, {Barnes}, \&
  {Prsa}}]{borucki2010a}
{Borucki}, W.~J., {Koch}, D., {Basri}, G., {et~al.} 2010, Science, 327, 977,
  \dodoi{10.1126/science.1185402}

\bibitem[{{Bryson} {et~al.}(2021){Bryson}, {Kunimoto}, {Kopparapu}, {Coughlin},
  {Borucki}, {Koch}, {Aguirre}, {Allen}, {Barentsen}, {Batalha}, {Berger},
  {Boss}, {Buchhave}, {Burke}, {Caldwell}, {Campbell}, {Catanzarite},
  {Chandrasekaran}, {Chaplin}, {Christiansen}, {Christensen-Dalsgaard},
  {Ciardi}, {Clarke}, {Cochran}, {Dotson}, {Doyle}, {Duarte}, {Dunham},
  {Dupree}, {Endl}, {Fanson}, {Ford}, {Fujieh}, {Gautier}, {Geary},
  {Gilliland}, {Girouard}, {Gould}, {Haas}, {Henze}, {Holman}, {Howard},
  {Howell}, {Huber}, {Hunter}, {Jenkins}, {Kjeldsen}, {Kolodziejczak},
  {Larson}, {Latham}, {Li}, {Mathur}, {Meibom}, {Middour}, {Morris}, {Morton},
  {Mullally}, {Mullally}, {Pletcher}, {Prsa}, {Quinn}, {Quintana}, {Ragozzine},
  {Ramirez}, {Sanderfer}, {Sasselov}, {Seader}, {Shabram}, {Shporer}, {Smith},
  {Steffen}, {Still}, {Torres}, {Troeltzsch}, {Twicken}, {Uddin}, {Van Cleve},
  {Voss}, {Weiss}, {Welsh}, {Wohler}, \& {Zamudio}}]{bryson2021}
{Bryson}, S., {Kunimoto}, M., {Kopparapu}, R.~K., {et~al.} 2021, \aj, 161, 36,
  \dodoi{10.3847/1538-3881/abc418}

\bibitem[{{Buchhave} {et~al.}(2014){Buchhave}, {Bizzarro}, {Latham},
  {Sasselov}, {Cochran}, {Endl}, {Isaacson}, {Juncher}, \&
  {Marcy}}]{buchhave2014}
{Buchhave}, L.~A., {Bizzarro}, M., {Latham}, D.~W., {et~al.} 2014, \nat, 509,
  593, \dodoi{10.1038/nature13254}

\bibitem[{{Cascioli} {et~al.}(2021){Cascioli}, {Hensley}, {De Marchi},
  {Breuer}, {Durante}, {Racioppa}, {Iess}, {Mazarico}, \&
  {Smrekar}}]{cascioli2021}
{Cascioli}, G., {Hensley}, S., {De Marchi}, F., {et~al.} 2021, \psj, 2, 220,
  \dodoi{10.3847/PSJ/ac26c0}

\bibitem[{{Chen} \& {Kipping}(2017)}]{chen2017}
{Chen}, J., \& {Kipping}, D. 2017, \apj, 834, 17,
  \dodoi{10.3847/1538-4357/834/1/17}

\bibitem[{{Christiansen} {et~al.}(2025){Christiansen}, {McElroy}, {Harbut},
  {Ciardi}, {Crane}, {Good}, {Hardegree-Ullman}, {Kesseli}, {Lund}, {Lynn},
  {Muthiar}, {Nilsson}, {Oluyide}, {Papin}, {Rivera}, {Swain}, {Susemiehl},
  {Tam}, {van Eyken}, \& {Beichman}}]{christiansen2025}
{Christiansen}, J.~L., {McElroy}, D.~L., {Harbut}, M., {et~al.} 2025, \psj, 6,
  186, \dodoi{10.3847/PSJ/ade3c2}

\bibitem[{{Dressing} \& {Charbonneau}(2013)}]{dressing2013}
{Dressing}, C.~D., \& {Charbonneau}, D. 2013, \apj, 767, 95,
  \dodoi{10.1088/0004-637X/767/1/95}

\bibitem[{{Dressing} \& {Charbonneau}(2015)}]{dressing2015b}
---. 2015, \apj, 807, 45, \dodoi{10.1088/0004-637X/807/1/45}

\bibitem[{{Ehrenreich} {et~al.}(2012){Ehrenreich}, {Vidal-Madjar}, {Widemann},
  {Gronoff}, {Tanga}, {Barth{\'e}lemy}, {Lilensten}, {Lecavelier Des Etangs},
  \& {Arnold}}]{ehrenreich2012a}
{Ehrenreich}, D., {Vidal-Madjar}, A., {Widemann}, T., {et~al.} 2012, \aap, 537,
  L2, \dodoi{10.1051/0004-6361/201118400}

\bibitem[{{Emsenhuber} {et~al.}(2021){Emsenhuber}, {Asphaug}, {Cambioni},
  {Gabriel}, \& {Schwartz}}]{emsenhuber2021a}
{Emsenhuber}, A., {Asphaug}, E., {Cambioni}, S., {Gabriel}, T. S.~J., \&
  {Schwartz}, S.~R. 2021, \psj, 2, 199, \dodoi{10.3847/PSJ/ac19b1}

\bibitem[{{Fressin} {et~al.}(2013){Fressin}, {Torres}, {Charbonneau}, {Bryson},
  {Christiansen}, {Dressing}, {Jenkins}, {Walkowicz}, \&
  {Batalha}}]{fressin2013}
{Fressin}, F., {Torres}, G., {Charbonneau}, D., {et~al.} 2013, \apj, 766, 81,
  \dodoi{10.1088/0004-637X/766/2/81}

\bibitem[{{Fulton} {et~al.}(2017){Fulton}, {Petigura}, {Howard}, {Isaacson},
  {Marcy}, {Cargile}, {Hebb}, {Weiss}, {Johnson}, {Morton}, {Sinukoff},
  {Crossfield}, \& {Hirsch}}]{fulton2017}
{Fulton}, B.~J., {Petigura}, E.~A., {Howard}, A.~W., {et~al.} 2017, \aj, 154,
  109, \dodoi{10.3847/1538-3881/aa80eb}

\bibitem[{{Garvin} {et~al.}(2022){Garvin}, {Getty}, {Arney}, {Johnson},
  {Kohler}, {Schwer}, {Sekerak}, {Bartels}, {Saylor}, {Elliott}, {Goodloe},
  {Garrison}, {Cottini}, {Izenberg}, {Lorenz}, {Malespin}, {Ravine}, {Webster},
  {Atkinson}, {Aslam}, {Atreya}, {Bos}, {Brinckerhoff}, {Campbell}, {Crisp},
  {Filiberto}, {Forget}, {Gilmore}, {Gorius}, {Grinspoon}, {Hofmann}, {Kane},
  {Kiefer}, {Lebonnois}, {Mahaffy}, {Pavlov}, {Trainer}, {Zahnle}, \&
  {Zolotov}}]{garvin2022}
{Garvin}, J.~B., {Getty}, S.~A., {Arney}, G.~N., {et~al.} 2022, \psj, 3, 117,
  \dodoi{10.3847/PSJ/ac63c2}

\bibitem[{{Goldblatt} {et~al.}(2013){Goldblatt}, {Robinson}, {Zahnle}, \&
  {Crisp}}]{goldblatt2013}
{Goldblatt}, C., {Robinson}, T.~D., {Zahnle}, K.~J., \& {Crisp}, D. 2013,
  Nature Geoscience, 6, 661, \dodoi{10.1038/ngeo1892}

\bibitem[{{Greene} {et~al.}(2023){Greene}, {Bell}, {Ducrot}, {Dyrek}, {Lagage},
  \& {Fortney}}]{greene2023}
{Greene}, T.~P., {Bell}, T.~J., {Ducrot}, E., {et~al.} 2023, \nat, 618, 39,
  \dodoi{10.1038/s41586-023-05951-7}

\bibitem[{{Heller} {et~al.}(2022){Heller}, {Harre}, \& {Samadi}}]{heller2022b}
{Heller}, R., {Harre}, J.-V., \& {Samadi}, R. 2022, \aap, 665, A11,
  \dodoi{10.1051/0004-6361/202141640}

\bibitem[{{Hill} {et~al.}(2023){Hill}, {Bott}, {Dalba}, {Fetherolf}, {Kane},
  {Kopparapu}, {Li}, \& {Ostberg}}]{hill2023}
{Hill}, M.~L., {Bott}, K., {Dalba}, P.~A., {et~al.} 2023, \aj, 165, 34,
  \dodoi{10.3847/1538-3881/aca1c0}

\bibitem[{{Hill} {et~al.}(2018){Hill}, {Kane}, {Seperuelo Duarte}, {Kopparapu},
  {Gelino}, \& {Wittenmyer}}]{hill2018}
{Hill}, M.~L., {Kane}, S.~R., {Seperuelo Duarte}, E., {et~al.} 2018, \apj, 860,
  67, \dodoi{10.3847/1538-4357/aac384}

\bibitem[{{Horner} {et~al.}(2020){Horner}, {Kane}, {Marshall}, {Dalba}, {Holt},
  {Wood}, {Maynard-Casely}, {Wittenmyer}, {Lykawka}, {Hill}, {Salmeron},
  {Bailey}, {L{\"o}hne}, {Agnew}, {Carter}, \& {Tylor}}]{horner2020b}
{Horner}, J., {Kane}, S.~R., {Marshall}, J.~P., {et~al.} 2020, \pasp, 132,
  102001, \dodoi{10.1088/1538-3873/ab8eb9}

\bibitem[{{Howell} {et~al.}(2014){Howell}, {Sobeck}, {Haas}, {Still},
  {Barclay}, {Mullally}, {Troeltzsch}, {Aigrain}, {Bryson}, {Caldwell},
  {Chaplin}, {Cochran}, {Huber}, {Marcy}, {Miglio}, {Najita}, {Smith},
  {Twicken}, \& {Fortney}}]{howell2014}
{Howell}, S.~B., {Sobeck}, C., {Haas}, M., {et~al.} 2014, \pasp, 126, 398,
  \dodoi{10.1086/676406}

\bibitem[{{Hsu} {et~al.}(2019){Hsu}, {Ford}, {Ragozzine}, \&
  {Ashby}}]{hsu2019c}
{Hsu}, D.~C., {Ford}, E.~B., {Ragozzine}, D., \& {Ashby}, K. 2019, \aj, 158,
  109, \dodoi{10.3847/1538-3881/ab31ab}

\bibitem[{{Ingersoll}(1969)}]{ingersoll1969c}
{Ingersoll}, A.~P. 1969, Journal of Atmospheric Sciences, 26, 1191,
  \dodoi{10.1175/1520-0469(1969)026<1191:TRGAHO>2.0.CO;2}

\bibitem[{{Kane} \& {Byrne}(2024)}]{kane2024b}
{Kane}, S.~R., \& {Byrne}, P.~K. 2024, Nature Astronomy, 8, 417,
  \dodoi{10.1038/s41550-024-02228-5}

\bibitem[{{Kane} \& {Gelino}(2012)}]{kane2012a}
{Kane}, S.~R., \& {Gelino}, D.~M. 2012, \pasp, 124, 323, \dodoi{10.1086/665271}

\bibitem[{{Kane} {et~al.}(2014){Kane}, {Kopparapu}, \&
  {Domagal-Goldman}}]{kane2014e}
{Kane}, S.~R., {Kopparapu}, R.~K., \& {Domagal-Goldman}, S.~D. 2014, \apjl,
  794, L5, \dodoi{10.1088/2041-8205/794/1/L5}

\bibitem[{{Kane} {et~al.}(2016){Kane}, {Hill}, {Kasting}, {Kopparapu},
  {Quintana}, {Barclay}, {Batalha}, {Borucki}, {Ciardi}, {Haghighipour},
  {Hinkel}, {Kaltenegger}, {Selsis}, \& {Torres}}]{kane2016c}
{Kane}, S.~R., {Hill}, M.~L., {Kasting}, J.~F., {et~al.} 2016, \apj, 830, 1,
  \dodoi{10.3847/0004-637X/830/1/1}

\bibitem[{{Kane} {et~al.}(2019){Kane}, {Arney}, {Crisp}, {Domagal-Goldman},
  {Glaze}, {Goldblatt}, {Grinspoon}, {Head}, {Lenardic}, {Unterborn}, {Way}, \&
  {Zahnle}}]{kane2019d}
{Kane}, S.~R., {Arney}, G., {Crisp}, D., {et~al.} 2019, Journal of Geophysical
  Research (Planets), 124, 2015, \dodoi{10.1029/2019JE005939}

\bibitem[{{Kane} {et~al.}(2021){Kane}, {Arney}, {Byrne}, {Dalba}, {Desch},
  {Horner}, {Izenberg}, {Mandt}, {Meadows}, \& {Quick}}]{kane2021d}
{Kane}, S.~R., {Arney}, G.~N., {Byrne}, P.~K., {et~al.} 2021, Journal of
  Geophysical Research (Planets), 126, e06643, \dodoi{10.1002/jgre.v126.2}

\bibitem[{{Kane} {et~al.}(2026){Kane}, {Bott}, {Goodis Gordon}, {Miles},
  {Ostberg}, {Byrne}, {Carone}, {Daylan}, {Garc{\'\i}a Mu{\~n}oz}, {Harada},
  {Hu}, {Izenberg}, {Kohler}, {Rice}, {Sagynbayeva}, {Scherf}, {Schwieterman},
  \& {Woitke}}]{kane2026a}
{Kane}, S.~R., {Bott}, K.~M., {Goodis Gordon}, K.~E., {et~al.} 2026, \pasp,
  138, 024404, \dodoi{10.1088/1538-3873/ae417d}

\bibitem[{{Kasting}(1988)}]{kasting1988c}
{Kasting}, J.~F. 1988, \icarus, 74, 472, \dodoi{10.1016/0019-1035(88)90116-9}

\bibitem[{{Kasting} {et~al.}(1993){Kasting}, {Whitmire}, \&
  {Reynolds}}]{kasting1993a}
{Kasting}, J.~F., {Whitmire}, D.~P., \& {Reynolds}, R.~T. 1993, \icarus, 101,
  108, \dodoi{10.1006/icar.1993.1010}

\bibitem[{{Kopparapu} {et~al.}(2014){Kopparapu}, {Ramirez}, {SchottelKotte},
  {Kasting}, {Domagal-Goldman}, \& {Eymet}}]{kopparapu2014}
{Kopparapu}, R.~K., {Ramirez}, R.~M., {SchottelKotte}, J., {et~al.} 2014, \apj,
  787, L29, \dodoi{10.1088/2041-8205/787/2/L29}

\bibitem[{{Kopparapu} {et~al.}(2013){Kopparapu}, {Ramirez}, {Kasting}, {Eymet},
  {Robinson}, {Mahadevan}, {Terrien}, {Domagal-Goldman}, {Meadows}, \&
  {Deshpande}}]{kopparapu2013a}
{Kopparapu}, R.~K., {Ramirez}, R., {Kasting}, J.~F., {et~al.} 2013, \apj, 765,
  131, \dodoi{10.1088/0004-637X/765/2/131}

\bibitem[{{Kunimoto} \& {Matthews}(2020)}]{kunimoto2020b}
{Kunimoto}, M., \& {Matthews}, J.~M. 2020, \aj, 159, 248,
  \dodoi{10.3847/1538-3881/ab88b0}

\bibitem[{{Leconte} {et~al.}(2013){Leconte}, {Forget}, {Charnay}, {Wordsworth},
  \& {Pottier}}]{leconte2013c}
{Leconte}, J., {Forget}, F., {Charnay}, B., {Wordsworth}, R., \& {Pottier}, A.
  2013, \nat, 504, 268, \dodoi{10.1038/nature12827}

\bibitem[{{Limaye} \& {Garvin}(2023)}]{limaye2023}
{Limaye}, S.~S., \& {Garvin}, J.~B. 2023, Frontiers in Astronomy and Space
  Sciences, 10, 1188096, \dodoi{10.3389/fspas.2023.1188096}

\bibitem[{{Luger} \& {Barnes}(2015)}]{luger2015b}
{Luger}, R., \& {Barnes}, R. 2015, Astrobiology, 15, 119,
  \dodoi{10.1089/ast.2014.1231}

\bibitem[{{Lustig-Yaeger} {et~al.}(2019){Lustig-Yaeger}, {Meadows}, \&
  {Lincowski}}]{lustigyaeger2019b}
{Lustig-Yaeger}, J., {Meadows}, V.~S., \& {Lincowski}, A.~P. 2019, \apjl, 887,
  L11, \dodoi{10.3847/2041-8213/ab5965}

\bibitem[{{Matuszewski} {et~al.}(2023){Matuszewski}, {Nettelmann}, {Cabrera},
  {B{\"o}rner}, \& {Rauer}}]{matuszewski2023}
{Matuszewski}, F., {Nettelmann}, N., {Cabrera}, J., {B{\"o}rner}, A., \&
  {Rauer}, H. 2023, \aap, 677, A133, \dodoi{10.1051/0004-6361/202245287}

\bibitem[{{Montalto} {et~al.}(2021){Montalto}, {Piotto}, {Marrese},
  {Nascimbeni}, {Prisinzano}, {Granata}, {Marinoni}, {Desidera}, {Ortolani},
  {Aerts}, {Alei}, {Altavilla}, {Benatti}, {B{\"o}rner}, {Cabrera}, {Claudi},
  {Deleuil}, {Fabrizio}, {Gizon}, {Goupil}, {Heras}, {Magrin}, {Malavolta},
  {Mas-Hesse}, {Pagano}, {Paproth}, {Pertenais}, {Pollacco}, {Ragazzoni},
  {Ramsay}, {Rauer}, \& {Udry}}]{montalto2021}
{Montalto}, M., {Piotto}, G., {Marrese}, P.~M., {et~al.} 2021, \aap, 653, A98,
  \dodoi{10.1051/0004-6361/202140717}

\bibitem[{{Nascimbeni} {et~al.}(2022){Nascimbeni}, {Piotto}, {B{\"o}rner},
  {Montalto}, {Marrese}, {Cabrera}, {Marinoni}, {Aerts}, {Altavilla},
  {Benatti}, {Claudi}, {Deleuil}, {Desidera}, {Fabrizio}, {Gizon}, {Goupil},
  {Granata}, {Heras}, {Magrin}, {Malavolta}, {Mas-Hesse}, {Ortolani}, {Pagano},
  {Pollacco}, {Prisinzano}, {Ragazzoni}, {Ramsay}, {Rauer}, \&
  {Udry}}]{nascimbeni2022}
{Nascimbeni}, V., {Piotto}, G., {B{\"o}rner}, A., {et~al.} 2022, \aap, 658,
  A31, \dodoi{10.1051/0004-6361/202142256}

\bibitem[{{Nascimbeni} {et~al.}(2025){Nascimbeni}, {Piotto}, {Cabrera},
  {Montalto}, {Marinoni}, {Marrese}, {Aerts}, {Altavilla}, {Benatti},
  {B{\"o}rner}, {Deleuil}, {Desidera}, {Gizon}, {Goupil}, {Granata}, {Heras},
  {Magrin}, {Malavolta}, {Mas-Hesse}, {Osborn}, {Pagano}, {Paproth},
  {Pollacco}, {Prisinzano}, {Ragazzoni}, {Ramsay}, {Rauer}, {Tkachenko}, \&
  {Udry}}]{nascimbeni2025}
{Nascimbeni}, V., {Piotto}, G., {Cabrera}, J., {et~al.} 2025, \aap, 694, A313,
  \dodoi{10.1051/0004-6361/202452325}

\bibitem[{{Ostberg} {et~al.}(2023{\natexlab{a}}){Ostberg}, {Kane}, {Lincowski},
  \& {Dalba}}]{ostberg2023c}
{Ostberg}, C., {Kane}, S.~R., {Lincowski}, A.~P., \& {Dalba}, P.~A.
  2023{\natexlab{a}}, \aj, 166, 213, \dodoi{10.3847/1538-3881/acfed2}

\bibitem[{{Ostberg} {et~al.}(2023{\natexlab{b}}){Ostberg}, {Kane}, {Li},
  {Schwieterman}, {Hill}, {Bott}, {Dalba}, {Fetherolf}, {Head}, \&
  {Unterborn}}]{ostberg2023a}
{Ostberg}, C., {Kane}, S.~R., {Li}, Z., {et~al.} 2023{\natexlab{b}}, \aj, 165,
  168, \dodoi{10.3847/1538-3881/acbfaf}

\bibitem[{{Pecaut} \& {Mamajek}(2013)}]{pecaut2013}
{Pecaut}, M.~J., \& {Mamajek}, E.~E. 2013, \apjs, 208, 9,
  \dodoi{10.1088/0067-0049/208/1/9}

\bibitem[{{Pertenais} {et~al.}(2021){Pertenais}, {Cabrera}, {Paproth},
  {Boerner}, {Grie{\ss}bach}, {Mogulsky}, \& {Rauer}}]{pertenais2021b}
{Pertenais}, M., {Cabrera}, J., {Paproth}, C., {et~al.} 2021, in Society of
  Photo-Optical Instrumentation Engineers (SPIE) Conference Series, Vol. 11852,
  International Conference on Space Optics {\textemdash} ICSO 2020, ed.
  B.~{Cugny}, Z.~{Sodnik}, \& N.~{Karafolas}, 118524Y,
  \dodoi{10.1117/12.2599820}

\bibitem[{{Petigura} {et~al.}(2013){Petigura}, {Marcy}, \&
  {Howard}}]{petigura2013a}
{Petigura}, E.~A., {Marcy}, G.~W., \& {Howard}, A.~W. 2013, \apj, 770, 69,
  \dodoi{10.1088/0004-637X/770/1/69}

\bibitem[{{Quanz} {et~al.}(2022){Quanz}, {Ottiger}, {Fontanet}, {Kammerer},
  {Menti}, {Dannert}, {Gheorghe}, {Absil}, {Airapetian}, {Alei}, {Allart},
  {Angerhausen}, {Blumenthal}, {Buchhave}, {Cabrera},
  {Carri{\'o}n-Gonz{\'a}lez}, {Chauvin}, {Danchi}, {Dandumont}, {Defr{\'e}re},
  {Dorn}, {Ehrenreich}, {Ertel}, {Fridlund}, {Garc{\'\i}a Mu{\~n}oz},
  {Gasc{\'o}n}, {Girard}, {Glauser}, {Grenfell}, {Guidi}, {Hagelberg},
  {Helled}, {Ireland}, {Janson}, {Kopparapu}, {Korth}, {Kozakis}, {Kraus},
  {L{\'e}ger}, {Leedj{\"a}rv}, {Lichtenberg}, {Lillo-Box}, {Linz}, {Liseau},
  {Loicq}, {Mahendra}, {Malbet}, {Mathew}, {Mennesson}, {Meyer}, {Mishra},
  {Molaverdikhani}, {Noack}, {Oza}, {Pall{\'e}}, {Parviainen}, {Quirrenbach},
  {Rauer}, {Ribas}, {Rice}, {Romagnolo}, {Rugheimer}, {Schwieterman},
  {Serabyn}, {Sharma}, {Stassun}, {Szul{\'a}gyi}, {Wang}, {Wunderlich},
  {Wyatt}, \& {LIFE Collaboration}}]{quanz2022a}
{Quanz}, S.~P., {Ottiger}, M., {Fontanet}, E., {et~al.} 2022, \aap, 664, A21,
  \dodoi{10.1051/0004-6361/202140366}

\bibitem[{{Ragazzoni} {et~al.}(2016){Ragazzoni}, {Magrin}, {Rauer}, {Pagano},
  {Nascimbeni}, {Piotto}, {Piazza}, {Levacher}, {Schweitzer}, {Basso}, {Bandy},
  {Benz}, {Bergomi}, {Biondi}, {Boerner}, {Borsa}, {Brandeker}, {Br{\"a}ndli},
  {Bruno}, {Cabrera}, {Chinellato}, {De Roche}, {Dima}, {Erikson}, {Farinato},
  {Munari}, {Ghigo}, {Greggio}, {Gullieuszik}, {Klebor}, {Marafatto},
  {Mogulsky}, {Peter}, {Rieder}, {Sicilia}, {Spiga}, {Viotto}, {Wieser},
  {Heras}, {Gondoin}, {Bodin}, \& {Catala}}]{ragazzoni2016a}
{Ragazzoni}, R., {Magrin}, D., {Rauer}, H., {et~al.} 2016, in Society of
  Photo-Optical Instrumentation Engineers (SPIE) Conference Series, Vol. 9904,
  Space Telescopes and Instrumentation 2016: Optical, Infrared, and Millimeter
  Wave, ed. H.~A. {MacEwen}, G.~G. {Fazio}, M.~{Lystrup}, N.~{Batalha},
  N.~{Siegler}, \& E.~C. {Tong}, 990428, \dodoi{10.1117/12.2236094}

\bibitem[{{Rauer} {et~al.}(2014){Rauer}, {Catala}, {Aerts}, {Appourchaux},
  {Benz}, {Brandeker}, {Christensen-Dalsgaard}, {Deleuil}, {Gizon}, {Goupil},
  {G{\"u}del}, {Janot-Pacheco}, {Mas-Hesse}, {Pagano}, {Piotto}, {Pollacco},
  {Santos}, {Smith}, {Su{\'a}rez}, {Szab{\'o}}, {Udry}, {Adibekyan}, {Alibert},
  {Almenara}, {Amaro-Seoane}, {Eiff}, {Asplund}, {Antonello}, {Barnes},
  {Baudin}, {Belkacem}, {Bergemann}, {Bihain}, {Birch}, {Bonfils}, {Boisse},
  {Bonomo}, {Borsa}, {Brand{\~a}o}, {Brocato}, {Brun}, {Burleigh}, {Burston},
  {Cabrera}, {Cassisi}, {Chaplin}, {Charpinet}, {Chiappini}, {Church},
  {Csizmadia}, {Cunha}, {Damasso}, {Davies}, {Deeg}, {D{\'\i}az}, {Dreizler},
  {Dreyer}, {Eggenberger}, {Ehrenreich}, {Eigm{\"u}ller}, {Erikson}, {Farmer},
  {Feltzing}, {de Oliveira Fialho}, {Figueira}, {Forveille}, {Fridlund},
  {Garc{\'\i}a}, {Giommi}, {Giuffrida}, {Godolt}, {Gomes da Silva}, {Granzer},
  {Grenfell}, {Grotsch-Noels}, {G{\"u}nther}, {Haswell}, {Hatzes},
  {H{\'e}brard}, {Hekker}, {Helled}, {Heng}, {Jenkins}, {Johansen},
  {Khodachenko}, {Kislyakova}, {Kley}, {Kolb}, {Krivova}, {Kupka}, {Lammer},
  {Lanza}, {Lebreton}, {Magrin}, {Marcos-Arenal}, {Marrese}, {Marques},
  {Martins}, {Mathis}, {Mathur}, {Messina}, {Miglio}, {Montalban}, {Montalto},
  {Monteiro}, {Moradi}, {Moravveji}, {Mordasini}, {Morel}, {Mortier},
  {Nascimbeni}, {Nelson}, {Nielsen}, {Noack}, {Norton}, {Ofir}, {Oshagh},
  {Ouazzani}, {P{\'a}pics}, {Parro}, {Petit}, {Plez}, {Poretti}, {Quirrenbach},
  {Ragazzoni}, {Raimondo}, {Rainer}, {Reese}, {Redmer}, {Reffert},
  {Rojas-Ayala}, {Roxburgh}, {Salmon}, {Santerne}, {Schneider}, {Schou},
  {Schuh}, {Schunker}, {Silva-Valio}, {Silvotti}, {Skillen}, {Snellen}, {Sohl},
  {Sousa}, {Sozzetti}, {Stello}, {Strassmeier}, {{\v{S}}vanda}, {Szab{\'o}},
  {Tkachenko}, {Valencia}, {Van Grootel}, {Vauclair}, {Ventura}, {Wagner},
  {Walton}, {Weingrill}, {Werner}, {Wheatley}, \& {Zwintz}}]{rauer2014}
{Rauer}, H., {Catala}, C., {Aerts}, C., {et~al.} 2014, Experimental Astronomy,
  38, 249, \dodoi{10.1007/s10686-014-9383-4}

\bibitem[{{Rauer} {et~al.}(2025){Rauer}, {Aerts}, {Cabrera}, {Deleuil},
  {Erikson}, {Gizon}, {Goupil}, {Heras}, {Walloschek}, {Lorenzo-Alvarez},
  {Marliani}, {Martin-Garcia}, {Mas-Hesse}, {O'Rourke}, {Osborn}, {Pagano},
  {Piotto}, {Pollacco}, {Ragazzoni}, {Ramsay}, {Udry}, {Appourchaux}, {Benz},
  {Brandeker}, {G{\"u}del}, {Janot-Pacheco}, {Kabath}, {Kjeldsen}, {Min},
  {Santos}, {Smith}, {Suarez}, {Werner}, {Aboudan}, {Abreu}, {Acu{\~n}a},
  {Adams}, {Adibekyan}, {Affer}, {Agneray}, {Agnor}, {Aguirre B{\o}rsen-Koch},
  {Ahmed}, {Aigrain}, {Al-Bahlawan}, {Alcacera Gil}, {Alei}, {Alencar},
  {Alexander}, {Alfonso-Garz{\'o}n}, {Alibert}, {Allende Prieto}, {Almeida},
  {Alonso Sobrino}, {Altavilla}, {Althaus}, {Alvarez Trujillo}, {Amarsi},
  {Ammler-von Eiff}, {Am{\^o}res}, {Andrade}, {Antoniadis-Karnavas},
  {Ant{\'o}nio}, {Aparicio del Moral}, {Appolloni}, {Arena}, {Armstrong},
  {Aroca Aliaga}, {Asplund}, {Audenaert}, {Auricchio}, {Avelino}, {Baeke},
  {Bailli{\'e}}, {Balado}, {Ballber Balaguer{\'o}}, {Balestra}, {Ball},
  {Ballans}, {Ballot}, {Barban}, {Barbary}, {Barbieri}, {Barcel{\'o} Forteza},
  {Barker}, {Barklem}, {Barnes}, {Barrado Navascues}, {Barragan}, {Baruteau},
  {Basu}, {Baudin}, {Baumeister}, {Bayliss}, {Bazot}, {Beck}, {Belkacem},
  {Bellinger}, {Benatti}, {Benomar}, {B{\'e}rard}, {Bergemann}, {Bergomi},
  {Bernardo}, {Biazzo}, {Bignamini}, {Bigot}, {Billot}, {Binet}, {Biondi},
  {Biondi}, {Birch}, {Bitsch}, {Bluhm Ceballos}, {B{\'o}di}, {Bogn{\'a}r},
  {Boisse}, {Bolmont}, {Bonanno}, {Bonavita}, {Bonfanti}, {Bonfils}, {Bonito},
  {Bonomo}, {B{\"o}rner}, {Boro Saikia}, {Borreguero Mart{\'\i}n}, {Borsa},
  {Borsato}, {Bossini}, {Bouchy}, {Bou{\'e}}, {Boufleur}, {Boumier},
  {Bourrier}, {Bowman}, {Bozzo}, {Bradley}, {Bray}, {Bressan}, {Breton},
  {Brienza}, {Brito}, {Brogi}, {Brown}, {Brown}, {Brun}, {Bruno}, {Bruns},
  {Buchhave}, {Bugnet}, {Buldgen}, {Burgess}, {Busatta}, {Busso}, {Buzasi},
  {Caballero}, {Cabral}, {Cabrero Gomez}, {Calderone}, {Cameron}, {Cameron},
  {Campante}, {Campos Gestal}, {Canto Martins}, {Cara}, {Carone}, {Carrasco},
  {Casagrande}, {Casewell}, {Cassisi}, {Castellani}, {Castro}, {Catala},
  {Catal{\'a}n Fern{\'a}ndez}, {Catelan}, {Cegla}, {Cerruti}, {Cessa},
  {Chadid}, {Chaplin}, {Charpinet}, {Chiappini}, {Chiarucci}, {Chiavassa},
  {Chinellato}, {Chirulli}, {Christensen-Dalsgaard}, {Church}, {Claret},
  {Clarke}, {Claudi}, {Clermont}, {Coelho}, {Coelho}, {Cogato}, {Colom{\'e}},
  {Condamin}, {Conde Garc{\'\i}a}, \& {Conseil}}]{rauer2025}
{Rauer}, H., {Aerts}, C., {Cabrera}, J., {et~al.} 2025, Experimental Astronomy,
  59, 26, \dodoi{10.1007/s10686-025-09985-9}

\bibitem[{{Ricker} {et~al.}(2015){Ricker}, {Winn}, {Vanderspek}, {Latham},
  {Bakos}, {Bean}, {Berta-Thompson}, {Brown}, {Buchhave}, {Butler}, {Butler},
  {Chaplin}, {Charbonneau}, {Christensen-Dalsgaard}, {Clampin}, {Deming},
  {Doty}, {De Lee}, {Dressing}, {Dunham}, {Endl}, {Fressin}, {Ge}, {Henning},
  {Holman}, {Howard}, {Ida}, {Jenkins}, {Jernigan}, {Johnson}, {Kaltenegger},
  {Kawai}, {Kjeldsen}, {Laughlin}, {Levine}, {Lin}, {Lissauer}, {MacQueen},
  {Marcy}, {McCullough}, {Morton}, {Narita}, {Paegert}, {Palle}, {Pepe},
  {Pepper}, {Quirrenbach}, {Rinehart}, {Sasselov}, {Sato}, {Seager},
  {Sozzetti}, {Stassun}, {Sullivan}, {Szentgyorgyi}, {Torres}, {Udry}, \&
  {Villasenor}}]{ricker2015}
{Ricker}, G.~R., {Winn}, J.~N., {Vanderspek}, R., {et~al.} 2015, Journal of
  Astronomical Telescopes, Instruments, and Systems, 1, 014003,
  \dodoi{10.1117/1.JATIS.1.1.014003}

\bibitem[{{Schlecker} {et~al.}(2024){Schlecker}, {Apai}, {Lichtenberg},
  {Bergsten}, {Salvador}, \& {Hardegree-Ullman}}]{schlecker2024}
{Schlecker}, M., {Apai}, D., {Lichtenberg}, T., {et~al.} 2024, \psj, 5, 3,
  \dodoi{10.3847/PSJ/acf57f}

\bibitem[{{Seager} {et~al.}(2022){Seager}, {Petkowski}, {Carr}, {Grinspoon},
  {Ehlmann}, {Saikia}, {Agrawal}, {Buchanan}, {Weber}, {French}, {Klupar},
  {Worden}, {Baumgardner}, \& {Venus Life Finder Mission Team}}]{seager2022a}
{Seager}, S., {Petkowski}, J.~J., {Carr}, C.~E., {et~al.} 2022, Aerospace, 9,
  385, \dodoi{10.3390/aerospace9070385}

\bibitem[{{Taylor} {et~al.}(2018){Taylor}, {Svedhem}, \& {Head}}]{taylor2018}
{Taylor}, F.~W., {Svedhem}, H., \& {Head}, J.~W. 2018, \ssr, 214, 35,
  \dodoi{10.1007/s11214-018-0467-8}

\bibitem[{{Turbet} {et~al.}(2021){Turbet}, {Bolmont}, {Chaverot}, {Ehrenreich},
  {Leconte}, \& {Marcq}}]{turbet2021}
{Turbet}, M., {Bolmont}, E., {Chaverot}, G., {et~al.} 2021, \nat, 598, 276,
  \dodoi{10.1038/s41586-021-03873-w}

\bibitem[{{Van Eylen} {et~al.}(2018){Van Eylen}, {Agentoft}, {Lundkvist},
  {Kjeldsen}, {Owen}, {Fulton}, {Petigura}, \& {Snellen}}]{vaneylen2018b}
{Van Eylen}, V., {Agentoft}, C., {Lundkvist}, M.~S., {et~al.} 2018, \mnras,
  479, 4786, \dodoi{10.1093/mnras/sty1783}

\bibitem[{{Way} \& {Del Genio}(2020)}]{way2020}
{Way}, M.~J., \& {Del Genio}, A.~D. 2020, Journal of Geophysical Research
  (Planets), 125, e06276, \dodoi{10.1029/2019JE006276}

\bibitem[{{Widemann} {et~al.}(2023){Widemann}, {Smrekar}, {Garvin},
  {Straume-Lindner}, {Ocampo}, {Schulte}, {Voirin}, {Hensley}, {Dyar},
  {Whitten}, {Nunes}, {Getty}, {Arney}, {Johnson}, {Kohler}, {Spohn},
  {O'Rourke}, {Wilson}, {Way}, {Ostberg}, {Westall}, {H{\"o}ning}, {Jacobson},
  {Salvador}, {Avice}, {Breuer}, {Carter}, {Gilmore}, {Ghail}, {Helbert},
  {Byrne}, {Santos}, {Herrick}, {Izenberg}, {Marcq}, {Rolf}, {Weller},
  {Gillmann}, {Korablev}, {Zelenyi}, {Zasova}, {Gorinov}, {Seth}, {Rao}, \&
  {Desai}}]{widemann2023}
{Widemann}, T., {Smrekar}, S.~E., {Garvin}, J.~B., {et~al.} 2023, \ssr, 219,
  56, \dodoi{10.1007/s11214-023-00992-w}

\bibitem[{{Wordsworth} \& {Pierrehumbert}(2013)}]{wordsworth2013b}
{Wordsworth}, R.~D., \& {Pierrehumbert}, R.~T. 2013, \apj, 778, 154,
  \dodoi{10.1088/0004-637X/778/2/154}

\bibitem[{{Zahnle} \& {Catling}(2017)}]{zahnle2017}
{Zahnle}, K.~J., \& {Catling}, D.~C. 2017, \apj, 843, 122,
  \dodoi{10.3847/1538-4357/aa7846}

\end{thebibliography}


\end{document}